\documentclass[twoside,preprintnumbers,amsmath,amssymb,pacs,nofootinbib]{revtex4}
\usepackage{graphicx,subfigure,hyperref}
\usepackage{dcolumn,marvosym}
\usepackage{braket}
\usepackage{float}
\usepackage{empheq,mathpazo}
\usepackage{color}
\usepackage{verbatim}

\newcommand{\p}{\partial}

\definecolor{Rood}{rgb}{1, 0, 0}

\begin{document}
\title{Holographic $\rho$ mesons in an external magnetic field}
\author{\bf N.~Callebaut}
\email{ncalleba.callebaut@ugent.be}
\affiliation{Ghent University, Department of Physics and Astronomy, Krijgslaan 281-S9, B-9000 Gent, Belgium}
\author{\bf D.~Dudal}
\email{david.dudal@ugent.be}
\affiliation{Ghent University, Department of Physics and Astronomy, Krijgslaan 281-S9, B-9000 Gent, Belgium}
\author{\bf H.~Verschelde}
\email{henri.verschelde@ugent.be,}
\affiliation{Ghent University, Department of Physics and Astronomy, Krijgslaan 281-S9, B-9000 Gent, Belgium}

\hyphenation{specifically}

\begin{abstract}
We study the $\rho$ meson in a uniform magnetic field $eB$ using a holographic QCD-model, more specifically a $\text{D}4/\text{D}8/\overline{\text{D}8}$ brane setup in the confinement phase at zero temperature with two quenched flavours. The parameters of the model are fixed by matching to corresponding dual field theory parameters at zero magnetic field. We show that the up- and down-flavour branes respond differently to the presence of the magnetic field in the dual QCD-like theory, as expected because of the different electromagnetic charge carried by up- and down-quark. We discuss how to recover the Landau levels, indicating an instability of the QCD vacuum at $eB = m_\rho^2$ towards a phase where charged $\rho$ mesons are condensed, as predicted by Chernodub using effective QCD-models. We improve on these existing effective QCD-model analyses by also taking into account the chiral magnetic catalysis effect, which tells us that the constituent quark masses rise with $eB$. This turns out to increase the value of the critical magnetic field for the onset of $\rho$ meson condensation to $eB \approx 1.1~m_\rho^2 \approx 0.67~\text{GeV}^2$. We briefly discuss the influence of pions, which turn out to be irrelevant for the condensation in the approximation made.
\end{abstract}
\maketitle

\section{Introduction}
Recently the interest in magnetically induced QCD effects has grown, as a magnetic field offers a controllable parameter that gives rise to new unexpected results that might lead to novel insights in QCD, e.g.~the chiral magnetic effect, which is related to CP-violating processes \cite{:2009uh}, chiral magnetic spiral/wave, a split between the deconfinement and chiral symmetry restoration temperature, etc. Since at the LHC very strong magnetic fields (of the order of $10^{15}$ Tesla) are expected to arise in heavy ion collisions \cite{Skokov:2009qp,Bzdak:2011yy,Deng:2012pc,Tuchin:2013ie}, this creates the perfect setting for experimental searches for possible interesting QCD effects in a strong magnetic field. This has resulted in a lot of activity in this relatively young research field. We refer to \cite{
Fukushima:2008xe,Kharzeev:2010gd, Basar:2010zd,Fukushima:2010fe,Fraga:2008qn,Mizher:2010zb,Mizher:2011wd,Gatto:2010pt,Gatto:2010qs,Chernodub:2011fr,Frasca:2011zn,Kashiwa:2011js,Nam:2011vn,Andersen:2011ip,Andersen:2012bq,Andersen:2012jf,Fukushima:2012xw,Fukushima:2012kc,Fraga:2012fs,Fraga:2012ev,Blaizot:2012sd,Simonov:2012if,Simonov:2012mf,Chernodub:2010qx,Chernodub:2011mc,Chernodub:2011gs,Hidaka:2012mz,Chernodub:2012zx,Chernodub:2012fi,Chernodub:2012mu,Buividovich:2008wf,Buividovich:2009wi,Buividovich:2010tn,Braguta:2011hq,D'Elia:2010nq,D'Elia:2011zu,Ilgenfritz:2012fw,Bali:2012zg,Bali:2011qj,Yamamoto:2011gk,Johnson:2008vna,Zayakin:2008cy,Yee:2009vw,Callebaut:2011uc,Preis:2010cq,Rebhan:2008ur,Gynther:2010ed,Bergman:2008sg,Ammon:2011je,Bu:2012mq,Filev:2011mt,Erdmenger:2011bw,Evans:2010xs} for a selection of relevant literature. An interesting review is \cite{boek}. During the cosmological electroweak phase transition, gigantic magnetic fields should have occurred as well \cite{Vachaspati:1991nm}. \\

Although realistic magnetic fields might have a short lifetime and/or be spacetime dependent, we will consider them fixed here, as is usually done.\\

One particular new QCD phenomenon, put forward by Chernodub in \cite{Chernodub:2010qx,Chernodub:2011mc} and further explored in \cite{Chernodub:2011gs,Chernodub:2012mu,Hidaka:2012mz,Chernodub:2012zx} is the possible instability\footnote{This was already mentioned earlier in \cite{Schramm:1991ex}.} of the QCD vacuum towards condensation of charged $\rho$ mesons when a sufficiently strong magnetic field is present, $eB_c \sim m_\rho^2$, somewhat analogous to a possible $W^\pm$-boson condensation in the electroweak model \cite{Ambjorn:1989sz,VanDoorsselaere:2012zb,Chernodub:2012fi}. In \cite{Chernodub:2010qx,Chernodub:2011mc} a phenomenological approach was adopted to argue the existence of a $\rho$ meson condensate, respectively using an effective quantum electrodynamics Lagrangian for the $\rho$ mesons and a NJL-model. Lattice evidence in favour of this effect appeared in \cite{Braguta:2011hq}. A consequence of such a condensation would be that the QCD vacuum at zero temperature becomes superconducting at sufficiently large magnetic field, forming a quite exotic phase. In addition, it should be a kind of anisotropic superconductor, see  \cite{Chernodub:2010qx,Chernodub:2011mc,Chernodub:2011gs} for more details on this. Quite recently, a similar instability was studied in a holographic toy model \cite{Ammon:2011je,Bu:2012mq}. Other examples of magnetic instabilities of bulk charged vectors in an AdS/CFT context can be found in \cite{Donos:2011pn,Almuhairi:2011ws}. 
\\

We set out to give further evidence for a magnetically induced $\rho$ meson condensation using a holographic approach. Some preliminary and incomplete results were already presented in \cite{Callebaut:2011uc}. The investigation of the condensation itself and ensuing conductivity properties of the vacuum will be relegated to future work. Here we intend to take the first step in this program, and show that a condensation should occur as we encounter a tachyonic instability in the $\rho$ meson sector. We shall rely on the much studied Sakai-Sugimoto model \cite{Sakai:2004cn,Sakai:2005yt}.\\

Holographic QCD-models give a description of hadronic physics through a dual supergravity theory in a higher-dimensional world. This working philosophy is based on the original Maldacena conjecture \cite{Maldacena:1997re}. The duality is valid for a large number of colours and at large 't Hooft coupling $\lambda$ where QCD itself is unmanageable, thus providing an analytical setting for studying non-perturbative QCD effects. The Sakai-Sugimoto model in particular \cite{Sakai:2004cn,Sakai:2005yt} uses a D4/D8/$\overline{\mbox{D8}}$ brane setup which manages to reproduce much of the low-energy physics of quenched QCD in the chiral limit, such as confinement and dynamical chiral symmetry breaking at low temperature, vector meson dominance, pion effective action, etc. A short review of the model is presented in Section \ref{Asect}.
We will discuss the effect of the magnetic field on the $\rho$ meson mass in both the original antipodal embedding considered by Sakai and Sugimoto (where the flavour branes merge at a value $u_0$ of the holographic radius equal to the cut-off of space $u_K$) and the more general non-antipodal embedding ($u_0 > u_K$). The latter enables us to take into account constituents of the $\rho$ meson, at least to some level. We extend the work of Sakai and Sugimoto on the numerical fixing of the holographic parameters for the $u_0 = u_K$ case to the case where $u_0 > u_K$ in Section \ref{B}. In order to get results in physical units we have fixed the free remaining parameters in the holographic model by matching them to phenomenological or experimental values for the constituent quark mass, the pion decay constant and the $\rho$ meson mass $m_\rho$, in absence of a magnetic field. Let us mention here already that throughout this article the notation $m_\rho$ will be solely used to denote the $\rho$ meson mass at \emph{zero} magnetic field. \\

In the Sakai-Sugimoto model at zero background magnetic field, the flavour $\text{D}8$- and $\overline{\text{D}8}$-branes form a stack of coinciding branes in the $\text{D}4$-brane background. In Section \ref{C}, we summarize how to introduce a background magnetic field $eB$ into the model. Only in the $u_0 > u_K$ case there is an effect of $eB$ on the embedding. Considering two flavours, we allow for the possibility that the up- and down-flavour branes respond to the magnetic field in a different way, which they turn out to do. This simply reflects in a geometrical way the up- and down-quarks coupling to the magnetic field with different electric charges. In other works like \cite{Johnson:2008vna,Yee:2009vw,Rebhan:2008ur,Preis:2010cq,Gynther:2010ed}, the influence of a magnetic field on some features of the Sakai-Sugimoto model was already investigated, however mostly for the $N_f=1$ case. The chiral magnetic effect was studied for example in \cite{Yee:2009vw,Gynther:2010ed}, whereas a part of the phase diagram in terms of a magnetic field and isospin chemical potential in \cite{Preis:2010cq}. Since we are interested in charged $\rho$ mesons, we need multiple flavours in our study, and thus have to extend on existing $N_f=1$ literature\footnote{In \cite{Rebhan:2008ur} the two-flavour Sakai-Sugimoto model was used. The embedding however was the same for up- and down-flavour branes, even in the presence of a magnetic background, because only the $u_0=u_K$ case was considered there.}. We find in Section \ref{D}, among other results, a heavier constituent mass for the up-quarks than for the down-quarks in the presence of a magnetic field (in the $u_0>u_K$ case).\\

In Section \ref{previous}, we derive the mass equation for the $\rho$ meson, recovering the Landau levels of \cite{Chernodub:2010qx} that indicate the instability of the vacuum, as the effective masses of the charged $\rho$ mesons' polarizations that are transverse w.r.t.~the applied magnetic field become imaginary  in the lowest Landau level when the magnetic field reaches the critical value $eB_c = m_\rho^2$.  An obvious drawback of the phenomenological model used in \cite{Chernodub:2010qx} is the fact that the $\rho$ mesons are regarded as point particles without internal structure. The magnetic field is so strong that it is not evident to trust that the underlying quark structure will be of no importance to the story.  Using a NJL model allows to consider internal quark dynamics \cite{Chernodub:2011mc} but the ensuing $\rho$ meson dynamics is then perhaps somewhat less phenomenologically valid than that of \cite{Chernodub:2010qx}.\\

In Section \ref{Landaunonantipodal} we take one effect of the constituents of the $\rho$ meson into account through the effect of chiral magnetic catalysis, which tells us that the constituent quarks get heavier in the presence of a background magnetic field, as will the $\rho$ meson itself (apart from the Landau shift lowering the effective mass). The resulting effective $\rho$ meson action resembles that of \cite{Chernodub:2010qx} but with an effective $eB$-dependent mass due to internal quark effects. This leads to a slightly larger value for the critical magnetic field, $eB_c \approx 1.1\,m_\rho^2$, at the onset of the predicted condensation. This increase is in qualitative agreement with the lattice data of \cite{Braguta:2011hq}. In Section \ref{CSpion} we discuss that pions, in a leading approximation in the inverse 't Hooft coupling, should not disturb the charged $\rho$ meson condensation, as the magnetic field induces a coupling only between the pions and non-transverse, w.r.t.~the magnetic field, $\rho$ meson polarizations. We summarize in Section \ref{conc} and point out how to improve our analysis in future work. In the Appendix, we have collected some technical details.

\section{Holographic setup}

\subsection{Review of the Sakai-Sugimoto model } \label{Asect}
The Sakai-Sugimoto model \cite{Sakai:2004cn,Sakai:2005yt} involves a system
of $N_f$ pairs of $\text{D}8-\overline{\text{D}8}$ flavour probe branes placed in a
D4-brane background
\begin{eqnarray}\label{backgr}
ds^2 &=& g_{mn} dx^m dx^n \quad(m,n = 0 \cdots 9) \nonumber \\
 &=& \left(\frac{u}{R}\right)^{3/2} (\eta_{\mu\nu}dx^\mu dx^\nu + f(u)d\tau^2) + \left(\frac{R}{u}\right)^{3/2}
\left( \frac{du^2}{f(u)} + u^2 d\Omega_4^2 \right), \nonumber\\
&&e^\phi = g_s \left(\frac{u}{R}\right)^{3/4} \hspace{2mm}, \quad
F_4 = \frac{N_c}{V_4}\epsilon_4 \hspace{2mm}, \quad f(u) = 1-\frac{u_K^3}{u^3}\,\,,
\end{eqnarray}
\noindent where $d\Omega_4^2$, $\epsilon_4$ and $V_4=8\pi^2/3$ are,
respectively, the line element, the volume form and the volume of a unit
four-sphere,  while $R$ is a constant parameter related to the string
coupling constant $g_s$, the number of colours $N_c$ and the string
length $\ell_s$ through $R^3 = \pi g_s N_c \ell_s^3$.
This background has a
natural cut-off at $u = u_K$ and the QCD-like theory is said to ``live'' at
$u \rightarrow \infty$. Imposing a smooth cut-off of space at $u = u_K$ determines in a unique way the period $\delta \tau$ of $\tau$:
\begin{eqnarray}
\delta \tau = \frac{4\pi}{3} \frac{R^{3/2}}{u_K^{1/2}} = 2\pi M_K^{-1}
\end{eqnarray}
with $M_K$ the inverse radius of the $\tau$-circle.
In this paper we will work with 3 colours, $N_c=3$, and 2 flavours, $N_f=2$.
We stress here that we ignore the back reaction of the flavour branes on the background, an assumption which is in principle only valid for $N_c \gg N_f$, which means that we are working in a holographic analogue of the quenched approximation. Unquenching the Sakai-Sugimoto model is a difficult task, see \cite{Burrington:2007qd}.

\begin{figure}[h!]
  \hfill
  \begin{minipage}[t]{.45\textwidth}
    \begin{center}
      \scalebox{0.7}{
  \includegraphics{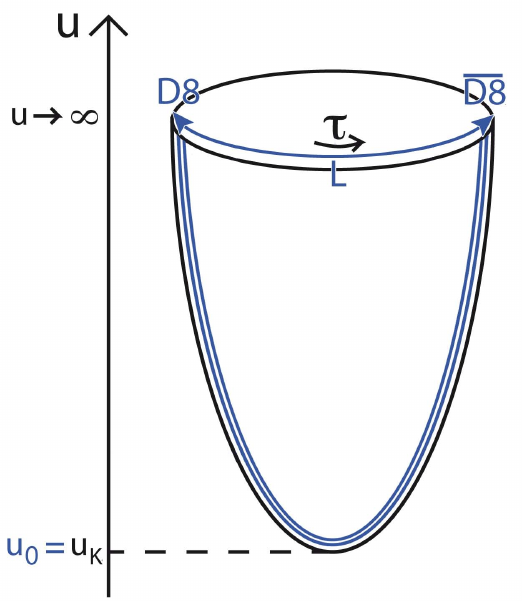}}
    \end{center}
  \end{minipage}
  \hfill
  \begin{minipage}[t]{.45\textwidth}
    \begin{center}
      \scalebox{0.7}{
  \includegraphics{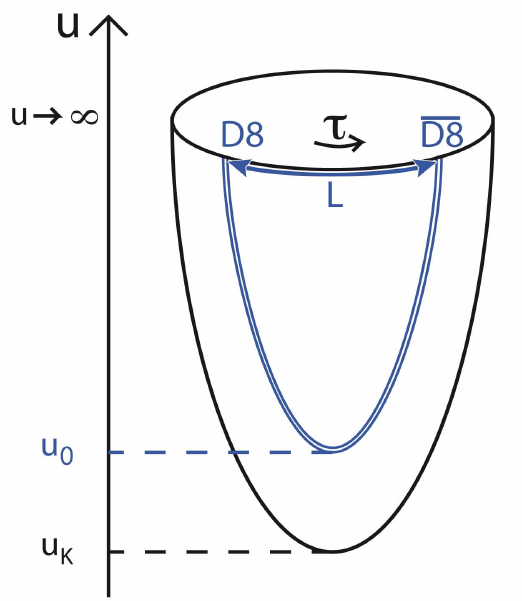}}
    \end{center}
  \end{minipage}
      \caption{The Sakai-Sugimoto model: antipodal ($u_0=u_K$) and non-antipodal ($u_0 > u_K$) embedding.}
	\label{SS}
  \hfill
\end{figure}

On the stack of $N_f$ coinciding $\text{D}8-\overline{\text{D}8}$ flavour pairs, there lives a $U(N_f)_L \times U(N_f)_R$ gauge theory, which is interpreted as corresponding to the global chiral symmetry in the dual QCD-like theory. The cigar-shape of the ($u,\tau$) subspace of the D4-brane background enforces a $\cup$-shaped embedding of the flavour branes, encoded in $u(\tau)$. This embedding represents the dynamical breaking of chiral symmetry $U(N_f)_L \times U(N_f)_R \rightarrow U(N_f)$ as the merging of the D8-branes and $\overline{\text{D}8}$-branes at $u = u_0$. The asymptotic separation $L$ (at $u\rightarrow \infty$) between D8- and $\overline{\mbox{D8}}$-branes, indicated in Figure \ref{SS}, is related to $u_0$ as
\begin{align}
L &= 2 \int_{u_0}^\infty \frac{du}{u'} \qquad (u' = du/d\tau)  \nonumber \\
   &= 2 \int_{u_0}^\infty du  \left(\frac{R}{u}\right)^{3/2} f^{-1} \sqrt{\frac{u_0^8 f_0}{u^8 f - u_0^8 f_0}}.
   \label{Lconf}
\end{align}
In the original setup of \cite{Sakai:2004cn,Sakai:2005yt} the flavour branes merge at the tip of the cigar, $u_0=u_K$, the embedding is antipodal. The more general non-antipodal embedding with $u_0 > u_K$ models non-zero
constituent quark mass \cite{Aharony:2006da}
\begin{equation}\label{constmass}
    m_q=\frac{1}{2\pi \alpha'}\int_{u_K}^{u_0}\frac{du}{\sqrt{f(u)}}\,\,,
\end{equation}
with $\alpha'= \ell_s^2$ the string tension. In this setup, unlike in the $u_0=u_K$ case, it is possible to model chiral magnetic catalysis \cite{Miransky:2002rp}: a magnetic field boosts the chiral symmetry breaking and hence the constituent quark masses. This phenomenon is by now generally accepted to occur in QCD, having received numerous confirmations  \cite{Fraga:2008qn,Mizher:2010zb,Mizher:2011wd,Gatto:2010pt,Gatto:2010qs,Nam:2011vn,Buividovich:2008wf,D'Elia:2011zu,Ilgenfritz:2012fw,Bali:2011qj,Bali:2012zg,Johnson:2008vna,Zayakin:2008cy,Shushpanov:1997sf}, albeit that the most realistic QCD lattice data display a richer behaviour than originally anticipated \cite{Bali:2011qj,Bali:2012zg}, probably related to quark mass/pion effects\footnote{In particular, regions of magnetic ``inhibition'' at finite temperature were reported, with analytical insights discussed in \cite{Fraga:2012fs,Fukushima:2012kc}.}. For the $N_f=1$ case of the Sakai-Sugimoto model, the catalysis was already observed in \cite{Johnson:2008vna}. The current quark masses are always zero in the Sakai-Sugimoto model, meaning that we are working in the chiral limit. This is closely related to the absence of a direct holographic description of the chiral condensate in the Sakai-Sugimoto model, which can only be overcome at the cost of great complication \cite{Bergman:2007pm}. \\ 

The $U(N_f)$ gauge field $A_m(x^\mu,u)$ ($m=0,1,2,3,u)$ living on the D8-branes
describes mesons in the boundary field theory. The action for this gauge
field is given by the non-Abelian DBI-action\footnote{The full non-Abelian generalization of the DBI-action is not known. The STr-prescription by Tseytlin which we will use is valid up to fourth order in the field strength, with deviations starting to appear at order $F^6$ \cite{Hashimoto:1997gm,Sevrin:2001ha}. The DBI action ignores derivative terms including $[F,F] \sim [D,D]F$ terms (ambiguity of the slow-varying fields approximation in non-Abelian case) \cite{Tseytlin:1997csa}. Because the 't Hooft coupling $\lambda$ is large in the validity regime of the sugra/QFT duality, we ignore in a first approximation the Chern-Simons part of the action in the analysis, being a factor $1/\lambda$ smaller than the DBI-action \cite{Sakai:2005yt,Aharony:2007uu}. We will come back to the discussion of this part of the action later.} \cite{Tseytlin:1997csa,Myers:2003bw,Howe:2006rv,Wulff:2007vj}    
\begin{equation} \label{nonabelian}
S_{DBI} = -T_8 \int d^4x \hspace{1mm} 2 \int_{u_0}^{\infty}  du \int \epsilon_4 \hspace{1mm} e^{-\phi}\hspace{1mm} \text{STr}
\sqrt{-\det \left[g_{mn}^{D8} + (2\pi\alpha') iF_{mn} \right]},
\end{equation}
where $T_8 = 1/((2\pi)^8 \ell_s^9)$ is the D8-brane tension, the factor 2 in front of the $u$-integration makes sure that we integrate over both halves of the $\cup$-shaped D8-branes,   $\text{STr}$ is the
symmetrized trace which is defined as
\begin{eqnarray} \text{STr}(F_1 \cdots F_n) = \frac{1}{n!} \text{Tr}(F_1 \cdots F_n + \mbox{all permutations}), \end{eqnarray}
$g_{mn}^{D8}$ the induced metric on the D8-branes, and
$F_{mn} = \partial_m A_n - \partial_n A_m + [A_m, A_n] = F_{mn}^a t^a$ the
field strength with anti-Hermitian generators
\begin{eqnarray} t^a  =  -\frac{i}{2} (\textbf{1}, \sigma_1, \sigma_2, \sigma_3), \quad
\text{Tr}(t^a t^b) = - \frac{\delta_{ab}}{2}, \quad [t^a,t^b] = \epsilon_{abc}t^c. \end{eqnarray}

The parameters $R, g_s , \ell_s, M_K$, $u_K$ and 't Hooft coupling $\lambda = g^2_{YM} N_c$ are
related through the following equations:
\begin{equation} \label{relaties}
R^3 = \frac{1}{2} \frac{\lambda \ell_s^2}{M_K},  \quad g_s =
\frac{1}{2\pi}\frac{g^2_{YM}}{M_K \ell_s},  \quad u_K = \frac{2}{9} \lambda
M_K \ell_s^2.
\end{equation}
Without loss of generality\footnote{All physical results are independent of the choice of $\lambda \ell_s^2$.} one can moreover set $\frac{2}{9} M_K^2 \ell_s^2 = \frac{1}{\lambda}$ \cite{Sakai:2005yt} which is the same as saying
\begin{equation} \label{SS2(2.4)a}
u_K = \frac{1}{M_K}.
\end{equation}
Consequently
\begin{eqnarray} R^3 = \frac{9}{4} \frac{1}{M_K^3} \quad \mbox{and}\quad \frac{1}{g_s \ell_s^3} = \frac{4\pi}{9} N_c M_K^3 = \frac{4\pi}{3} M_K^3. \end{eqnarray}
Following \cite{Sakai:2005yt}, we define
\begin{equation} \label{kappa}
\kappa = \frac{\lambda N_c}{216 \pi^3} = \frac{\lambda}{72 \pi^3}
\end{equation}
for notational convenience.

\subsection{Numerical fixing of the holographic parameters } \label{B}

In the model with antipodal embedding there are six parameters, $R, \kappa, \ell_s, M_K, g_s$ and $L$, related to each other through the four relations (\ref{relaties})-(\ref{SS2(2.4)a}). In \cite{Sakai:2005yt} the remaining independent parameters $M_K$ and $\kappa$ were fixed to GeV units by matching to the QCD input values
\begin{eqnarray} \quad f_\pi = 0.093  \mbox{ GeV} \quad \mbox{and} \quad m_\rho = 0.776 \mbox{ GeV}, \end{eqnarray}
resulting in
\begin{equation} \label{antipodalvalues}
M_K \approx 0.949 \mbox{  GeV} \quad
\mbox{and}\quad \kappa  = \frac{\lambda N_c}{216 \pi^3} \approx 0.00745.
\end{equation}

We extend this to the model with non-antipodal embedding where there are seven parameters, $R, \kappa, \ell_s, M_K, u_0, g_s$ and $L$.
These are completely determined by the four relations (\ref{relaties})-(\ref{SS2(2.4)a}), and the three extra requirements that the computable numerical values for the constituent quark mass $m_q$, the pion decay constant $f_\pi$ and the $\rho$ meson mass $m_\rho$, in absence of magnetic field, match to the phenomenologically or experimentally obtained QCD input values
\begin{eqnarray} m_q = 0.310 \mbox{ GeV}, \quad f_\pi = 0.093  \mbox{ GeV} \quad \mbox{and} \quad m_\rho = 0.776 \mbox{ GeV}. \end{eqnarray}
Adapting the analysis of Sakai and Sugimoto \cite{Sakai:2004cn} to the more general case $u_0 > u_K$, see also \cite{Peeters:2006iu,Peeters:2007ab}, we derive the mass eigenvalue equation for the vector meson sector and the expression for $f_\pi$ as  functions of
$M_K$, $u_0$ and $\kappa$, and thus indirectly (through the relation (\ref{Lconf})) as  functions of the three unknown parameters
$M_K$, $L$ and $\kappa$.\\

In the confinement phase, the non-Abelian DBI-action (\ref{nonabelian})
becomes, to second order in the field strength,
\begin{equation} \label{secondorder}
S_{DBI} = \widetilde V  \int d^4x \hspace{1mm}2 \int_{u_0}^\infty du  \hspace{1mm}\text{Tr} \left\{ u^{-1/2} \gamma^{1/2} R^3 F_{\mu\nu}F^{\mu\nu} + 2u^{5/2} \gamma^{-1/2} \eta^{\mu\nu}F_{\mu u}F_{\nu u} \right\} + \mathcal O(F^4),
\end{equation}
with \begin{eqnarray} \widetilde V = T_8 V_4 g_s^{-1} R^{3/2} \frac{1}{4} (2\pi\alpha')^2 \end{eqnarray}
and \begin{eqnarray} \gamma(u) = \frac{u^8}{f(u) u^8 - f(u_0) u_0^8}, \end{eqnarray} \label{gamma}
and where the $\text{STr}$ is replaced by $\text{Tr}$ as
\begin{eqnarray} \text{STr}(t^a t^b) = \text{Tr} (t^a t^b).\end{eqnarray}

Assuming the flavour gauge field components $A_\mu(x^\mu,u)$ can be expanded in complete sets $\left\{\psi_n(u)\right\}_{n\geq 1}$,
\begin{eqnarray}
A_\mu(x,u) &=& \sum_{n \geq 1} B_\mu^{(n)}(x) \psi_n(u)\\
\Rightarrow F_{\mu\nu}F^{\mu\nu} &=& F_{\mu\nu}^{(m)}F^{(n)\mu\nu} \psi_m \psi_n \nonumber \\\text{with}~  F_{\mu\nu}^{(m)} &=& \partial_\mu B_\nu^{(m)} - \partial_\nu B_\mu^{(m)},  \quad
F_{\mu u}F_{\nu u} ~=~ B_\mu^{(m)}B_\nu^{(n)}\partial_u \psi_m \partial_u \psi_n + \cdots,\nonumber
\end{eqnarray}
the part of the action (\ref{secondorder}) in $B_\mu^{(n)}(x)$ reduces to the effective 4-dimensional action
\begin{eqnarray}
S = - \int d^4 x \left\{ \frac{1}{4} F_{\mu\nu}^{(n)a}F^{(n)\mu\nu a} + \frac{1}{2} m_n^2 B_\mu^{(n)a} B^{(n)\mu a} \right\},
\end{eqnarray}
describing vector mesons $B_\mu^{(n)}$ with masses $m_n$, if the $\psi_n(u)$ are subject to:
\begin{equation} \label{orthonormconditiepsi}
\widetilde V \int_{u_0}^{\infty} du \hspace{1mm} u^{-1/2} \gamma^{1/2} R^3 \psi_m \psi_n = \frac{1}{4} \delta_{mn},
\end{equation}
and
\begin{equation} \label{blablaalg}
2 \widetilde V \int_{u_0}^{\infty} du \hspace{1mm} u^{5/2} \gamma^{-1/2} (\partial_u \psi_m)(\partial_u \psi_n) = \frac{1}{2} m_n^2 \delta_{mn} \quad (m,n \geq 1).
\end{equation}
The conditions (\ref{orthonormconditiepsi}) and (\ref{blablaalg}) combine, using partial integration (and hereby implicitly assuming the $\psi_n$ to be normalizable functions), to the eigenvalue equation
\begin{equation} \label{finiteTeigwvgl}
-u^{1/2} \gamma^{-1/2} \partial_u (u^{5/2} \gamma^{-1/2} \partial_u \psi_n) = R^3 m_n^2 \psi_n.
\end{equation}
To include pions, following the original discussion in \cite{Sakai:2004cn}, it is convenient to introduce a new coordinate $z$ that is related to $u$ through
\begin{eqnarray} u^3 = u_0^3 + u_0 z^2, \end{eqnarray}
going from $-\infty$ to $+\infty$ along the flavour branes and thus allowing the description of both boundaries at $u\rightarrow \infty$ of the $\cup$-shaped flavour branes.
In this new coordinate $z$, the action (\ref{secondorder}) reads (denoting $u(z)$ as $u_z$ for readability)
\begin{equation}
S_{DBI} = \widetilde V \int d^4x  \int_{-\infty}^\infty dz \hspace{1mm}\text{Tr} \left\{
\frac{2}{3} u_0 u_z^{3/2} \gamma' R^3 F_{\mu\nu}F^{\mu\nu} +
\frac{3}{u_0}  \frac{u_z^{1/2}}{\gamma'} \eta^{\mu\nu}F_{\mu z}F_{\nu z}  \right\} + \mathcal O(F^4),
\end{equation}
with
\begin{eqnarray} \gamma'(z) = \frac{|z| \sqrt \gamma}{u_z^4} = \sqrt{\frac{z^2}{u_z^{5}(u_z^3 - u_K^3) - (u_0^8 - u_0^5 u_K^3)}}, \end{eqnarray}
and the condition (\ref{blablaalg})
\begin{equation} \label{blablaalgz}
\frac{\widetilde V}{2} \int_{-\infty}^{\infty} dz \hspace{1mm} \frac{3}{u_0} \frac{u_z^{1/2}}{\gamma'} (\partial_z \psi_m)(\partial_z \psi_n) = \frac{1}{2} m_n^2 \delta_{mn} \quad (m,n \geq 1).
\end{equation}
The flavour gauge field component\footnote{The gauge field components along the four-sphere are assumed to vanish, $A_\alpha$ = 0, and $A_\mu$ and $A_z$ are assumed to be independent of the four-sphere coordinates.} $A_z$  is expanded in the complete set $\left\{\phi_n(z)\right\}_{n\geq 0}$:
\begin{eqnarray} \label{gaugefieldexpansion}
A_\mu(x,z) &=& \sum_{n \geq 1} B_\mu^{(n)}(x) \psi_n(z), \quad A_z(x,z) = \sum_{n \geq 0} \phi^{(n)}(x) \phi_n(z)
\\
\Rightarrow F_{\mu\nu}F^{\mu\nu} &=& F_{\mu\nu}^{(m)}F^{(n)\mu\nu} \psi_m \psi_n\,,\quad F_{\mu z}F_{\nu z} = (\partial_\mu \phi^{(m)} \phi_m - B_\mu^{(m)}\partial_z \psi_m)(\partial_\nu \phi^{(n)} \phi_n - B_\nu^{(n)}\partial_z \psi_n).\nonumber
\end{eqnarray}
Demanding a canonical normalization of the kinetic term for the $\phi^{(n)}(x^\mu)$ fields in the effective 4-dimensional action then leads to the orthonormality condition
\begin{equation} \label{orthonormconditiephi0z}
\frac{\widetilde V}{2} \int_{-\infty}^{\infty} dz \hspace{1mm} \frac{3}{u_0} u_z^{1/2}\gamma'^{-1} \phi_m \phi_n = \frac{1}{2} \delta_{mn} \quad (m,n \geq 0).
\end{equation}
From the last condition (\ref{orthonormconditiephi0z}) and (\ref{blablaalgz}) it follows that
\begin{eqnarray} \phi_n = m_n^{-1} \partial_z \psi_n \quad \mbox{for $n\geq 1$}\end{eqnarray}
and since $\phi_0 \perp \partial_z \psi_n$ for all $n \geq 1$,
\begin{equation}
\frac{\widetilde V}{2} \int_{-\infty}^{\infty} dz \hspace{1mm}  \frac{3}{u_0} u_z^{1/2}\gamma'^{-1} \phi_0 \partial_z \psi_n = 0,
\end{equation}
we can set
\begin{eqnarray}
\phi_0 = c \gamma' u_z^{-1/2} = c \frac{\gamma'}{(u_0^3+u_0 z^2)^{1/6}}
\end{eqnarray}
with the normalization constant $c$ determined by
\begin{eqnarray} \frac{\widetilde V}{2} c^2 \int_{-\infty}^{\infty} dz \hspace{1mm} \frac{3}{u_0} u_z^{-1/2}\gamma' = \frac{1}{2}. \end{eqnarray}
Now $\psi_0$ is defined through $\phi_0 = \partial_z \psi_0$, and $\hat \psi_0$ as a multiple of $\psi_0$,
\begin{eqnarray} \hat \psi_0 = c' \int_0^z dz\frac{\gamma'}{(u_0^3+u_0 z^2)^{1/6}}, \end{eqnarray}
that fulfills
\begin{eqnarray} \hat \psi_0(\pm \infty) = \pm \frac{1}{2}. \end{eqnarray}
We can then rewrite the expansion for the gauge field as
\begin{eqnarray} A_\mu = \xi_+ \partial_\mu \xi_+^{-1} \psi_+ + \xi_- \partial_\mu \xi_-^{-1} \psi_- + \sum_n B_\mu^{(n)} \psi_n \,\, , \quad A_z=0 \end{eqnarray}
with
\begin{eqnarray} \psi_\pm (z) = \frac{1}{2} \pm \hat \psi_0 \quad \mbox{ such that } \psi_+(\infty) = \psi_-(-\infty) = 1 \mbox{ and } \psi_+(-\infty)= \psi_-(\infty)= 0\end{eqnarray}
and
\begin{eqnarray} \xi_\pm^{-1}(x^\mu) = \mathcal P \exp\left\{- \int_0^{\pm \infty} dz' A_z(x^\mu, z')\right\}.\end{eqnarray}
One stays in the $A_z = 0$ gauge under residual gauge transformations $g(x^\mu, z = 0) = h(x^\mu)$. Fixing the gauge to $\xi_-=1$, we have the following gauge field expansion \cite{Sakai:2004cn}
\begin{eqnarray} A_\mu(x,z) = U^{-1}(x)\partial_\mu U(x) \psi_+(z) + \sum_{n\geq1} B_\mu^{(n)}(x) \psi_n(z), \quad A_z = 0. \end{eqnarray}
The pion field is defined as
\begin{eqnarray} U(x^\mu) = \mathcal P \exp{ \left\{-\int_{-\infty}^{\infty} dz' A_z(x^\mu, z') \right\}} \end{eqnarray}
in order to use the same pion field as the one used in the sigma-model for low-energy effective QCD, $U(x^\mu)\equiv  e^{2 i \pi(x^\mu)/f_\pi}, \pi(x^\mu) = \pi_a t^a$,
where the kinetic part of the action for the mesons is given by
\begin{eqnarray} \int d^4 x \frac{f_\pi^2}{4} \hspace{1mm} \text{Tr} (U^\dagger \partial_\mu U)^2.  \end{eqnarray}
Equating this action with the corresponding term in the Sakai-Sugimoto action after plugging in the expansion for the gauge field,
\begin{eqnarray} \widetilde V \int d^4x  \int_{-\infty}^\infty dz \hspace{1mm}\text{Tr} \left\{
\frac{3}{u_0}  \frac{(u_0^3+u_0 z^2)^{1/6}}{\gamma'} \eta^{\mu\nu}F_{\mu z}F_{\nu z}  \right\} =
\widetilde V \int d^4x  \int_{-\infty}^\infty dz \hspace{1mm} \frac{3}{u_0} \frac{c}{\phi_0} (\partial_z \psi^+)^2  \hspace{1mm} \text{Tr}(U^\dagger \partial_\mu U)^2, \end{eqnarray}
leads to the identification
\begin{eqnarray} \frac{f_\pi^2}{4} = \widetilde V \frac{3}{u_0} c' = \widetilde V \frac{3}{u_0} \left( 2 \int_{0}^\infty dz \frac{\gamma'}{(u_0^3+u_0 z^2)^{1/6}} \right)^{-1}, \end{eqnarray}
or
\begin{eqnarray} f_\pi^2(M_K,u_0,\kappa) = \frac{4}{3} \kappa M_K^{7/2} \frac{3}{u_0}\left( 2 \int_{0}^\infty dz \frac{\gamma'}{(u_0^3+u_0 z^2)^{1/6}} \right)^{-1}, \end{eqnarray}
where we have used the relations (\ref{relaties}), (\ref{SS2(2.4)a}) and the definition (\ref{kappa}), to determine the volume factor $\widetilde V$ in the action:
\begin{eqnarray} \widetilde V = \frac{1}{3} \kappa M_K^{7/2}.  \end{eqnarray}

We now have all the ingredients to numerically fix the remaining parameters $M_K$, $u_0$ (thus $L$) and $\kappa$.
First, we determine $\kappa(M_K,u_0)$ by demanding the constituent quark mass to be 0.310 GeV,
\begin{equation}  \label{mq}
m_q(M_K,u_0,\kappa) =8\pi^2 M_K^2 \kappa \int_{1/M_K}^{u_0} du \frac{1}{\sqrt{1 - \frac{1}{(M_Ku)^3}}} = 0.310 \mbox{ GeV} \Longrightarrow \kappa(M_K,u_0).
\end{equation}
Then we use the experimental value for the pion decay constant to find $u_0(M_K)$,
\begin{eqnarray}
f_\pi(M_K,u_0,\kappa(M_K,u_0))  =  f_\pi(M_K,u_0) = 0.093 \mbox{ GeV} \Longrightarrow u_0(M_K).
\end{eqnarray}
Finally, we solve the eigenvalue equation (\ref{finiteTeigwvgl}), which is now a function of  $M_K$ only, for $m_{n=1}$. We refer the reader to the Appendix for more details.
The value of $M_K$ is then determined such that $m_{n=1} = m_\rho = 0.776$ GeV,
the $\rho$ meson being the lightest meson in the vector meson tower. One identifies  $B_\mu^{(n=1)a}$ with $\rho_\mu^{a}$ ($a=1,2$), which can be recombined into the charged $\rho_{\mu}^{\pm}$ mesons, $B_\mu^{(n=1)3}$ with the neutral $\rho_\mu^{0}$ meson, and $B_\mu^{(n=1)0}$ with the $\omega_\mu$ meson. From these identifications it follows that $m_\rho = m_\omega = m_{n=1}$ in the Sakai-Sugimoto model.\\

The results of our numerical analysis are
\begin{equation}
M_K \approx 0.7209 \mbox{  GeV}, \quad  \frac{u_0}{u_K} \approx 1.38 \quad
\mbox{and}\quad \kappa  = \frac{\lambda N_c}{216 \pi^3} \approx 0.006778,
\end{equation}
or
\begin{equation} \label{values}
M_K \approx 0.7209 \mbox{  GeV}, \quad L \approx  1.574 \mbox{  GeV}^{-1} \quad
\mbox{and}\quad \kappa  = \frac{\lambda N_c}{216 \pi^3} \approx 0.006778,
\end{equation}
where we used the formula (\ref{Lconf}) describing the one-to-one relation between $L$ and $u_0$.
The value found for $L$ is
approximately 2.8 times smaller than the maximum value of $L$, given by
\begin{eqnarray}
L_{max} = \frac{\delta \tau}{2} = \frac{\pi}{M_K} \approx 4.358 \mbox{  GeV}^{-1}. \label{Lmax}
\end{eqnarray}

From the values (\ref{values}) we do extract a relatively large 't Hooft coupling, $\lambda\approx 15$. This allows us to ignore the Chern-Simons part of the action, which is a factor $\lambda$ smaller than the DBI-part. Nevertheless we will briefly comment on the contributions to the $\rho$ meson mass equation originating from the Chern-Simons action in Section \ref{previous}. We also remark that for these values of the holographic parameters the numerical value for the effective string tension between a quark and an antiquark in this background, given by \cite{Kinar:1998vq}
\begin{equation}
\sigma = \frac{1}{2\pi\alpha'} \sqrt{-g_{00}(u_K) g_{11}(u_K)} \approx 0.19 \text{ GeV}^2,
\end{equation}
is in good accordance with the value calculated on the lattice for pure SU(3) QCD, $\sigma \approx 0.18$-$0.19 \text{ GeV}^2$, as reported in \cite{Bali:1992ru,Sommer:1993ce}. This is a nice illustration that the fixed values do have a reasonable QCD resemblance.

\subsection{Turning on a uniform magnetic field }\label{C}
Under gauge transformations $g\in U(N_f)$, the flavour gauge field transforms as
\begin{eqnarray} \label{gaugetransf}
A_m(x^m) \rightarrow g A_m(x^m) g^{-1} + g \partial_m g^{-1} \quad (m=\mu,z).
\end{eqnarray}
Since we have assumed the eigenfunctions $\psi_n (n\geq 1)$ to be normalizable ($\psi_n(z\rightarrow \pm \infty) = 0$), the expansion (\ref{gaugefieldexpansion}) implicitly assumes we are working in the gauge $A_\mu(z\rightarrow \pm \infty) = 0$. The gauge potential can be made to vanish asymptotically by applying a gauge transformation $g(x ^\mu,z) = U(x^\mu,z)$, that cancels the asymptotic pure gauge configuration
\begin{equation} \label{puregauge}
 A_m(x^\mu,z\rightarrow \pm \infty) = U_{\pm}^{-1}(x^\mu,z) \partial_m U_\pm(x^\mu,z)
\end{equation}
that ensures a finite effective four-dimensional action. For arbitrary $N_f>2$, the homotopy group for the functions $U:\mathbb R_4 \cup \infty \simeq S_4  \rightarrow U(N_f)$, $x^\mu \rightarrow U(x^\mu,z)$ is trivial, $\pi_4(U(N_f))=0$, so a
continuously interpolating $U(N_f)$-valued function $U(x^\mu,z)$ that fulfills $U(x^\mu,z\rightarrow \pm \infty)=U_\pm(x^\mu,z)$ can always be found. The case $N_f=2$, which we consider, is an exception since $\pi_4(U(2)) = \mathbb{Z}_2$. In the seminal paper of Sakai and Sugimoto \cite{Sakai:2004cn}, it was assumed that $N_f\neq2$. Yet it appears to be still possible to consider the gauge $A_m(z\rightarrow \pm \infty) = 0$. For $N_f=2$, there will exist a 2 by 2 matrix function $U(x^\mu,z)$ interpolating between $U_+$ and $U_-$ (if they are homotopic) \'or between $U_+$ and $\widetilde U_-$ (if $U_+$ and $U_-$ are not homotopic), with $\widetilde U_-$ defined as $\sigma_3 U_-$, so that  $\widetilde U_-$ is homotopic to the $\mathbb{Z}_2$ element $\mp 1$ if $U_-$ is homotopic to the $\mathbb{Z}_2$ element $\pm 1$. The role of the $\sigma_3$ multiplication is to switch sign of a row in $U_-$, and thus also of the determinant. The sign of the determinant determines whether a $U
 (2)$ matrix is homotopic to $-1$ or $+1$. Since the asymptotic pure gauge configuration (\ref{puregauge}) also equals $\widetilde U_{\pm}^{-1}(x^\mu,z) \partial_m \widetilde U_\pm(x^\mu,z)$, as can be easily verified, the gauge transformation $g(x ^\mu,z) = U(x^\mu,z)$ will again cancel the asymptotic gauge potential. This argument\footnote{We thank J.~Van Doorsselaere for discussion on this point.} extends the validity of the original Sakai-Sugimoto reasoning to the $N_f=2$ case.\\

One does not leave the gauge $A_m(z\rightarrow \pm \infty) = 0$  under gauge transformations $h\in U(N_f)$ that adopt $x^\mu$-independent boundary values $(h_+, h_-) = (\lim_{z \rightarrow + \infty} h, \lim_{z \rightarrow - \infty} h)$. These boundary values of the residual gauge symmetry transformation $h$ are interpreted as a global chiral symmetry transformation $(h_L,h_R) \in U(N_f)_L \times U(N_f)_R$ in the dual QCD-like theory. By  ``lightly'' gauging this chiral symmetry, i.e.~making $h_L=h_R=h(z\rightarrow \pm \infty)=h$ dependent on $x^\mu$, 
one leaves the $A_m(z\rightarrow \pm \infty) = 0$ gauge, the boundary value of the gauge field $A_m(z\rightarrow \pm \infty)$ to be interpreted as an external background vector field $\overline A_\mu$ in the boundary field theory coupling to the quarks through a covariant derivative $\mathcal D_\mu = \partial_\mu + \overline A_\mu$ such that the Dirac action $\overline{\psi} i \gamma_\mu \mathcal D_\mu \psi$ remains invariant under local $U(N_f)$ transformations. 
Throughout this article we will be working in the $A_z=0$ gauge instead, where we have the following expansion of the gauge field \cite{Sakai:2005yt} (with our $\overline A_\mu$ equal to $A_L = A_R = \mathcal V$ in the notation of \cite{Sakai:2005yt}) 
\begin{equation}
A_\mu = (\xi_+ \partial_\mu \xi_+^{-1} + \xi_+ \overline A_\mu \xi_+^{-1}) \psi_+ + (\xi_- \partial_\mu \xi_-^{-1} + \xi_- \overline A_\mu \xi_-^{-1}) \psi_- + \sum_n B_\mu^{(n)} \psi_n.
\end{equation}
Fixing the residual gauge symmetry within this gauge to $\xi_+^{-1}(x^\mu) = \xi_-(x^\mu)$, the expansion becomes
\begin{eqnarray} \label{expansionvectorpion}
A_\mu = \overline A_\mu + \frac{1}{2 f_\pi^2} [\pi, \partial_\mu \pi] + \frac{i}{f_\pi} \left(\partial_\mu \pi - [\pi,\overline A_\mu] \right) \psi_0 +  \sum_n B_\mu^{(n)} \psi_n. \end{eqnarray}
In this paper we shall not look at the pions in detail, being mainly interested in the $\rho$ mesons (identified with $B_\mu^{(1)}$), so essentially we will use the gauge field expansion
\begin{eqnarray} \label{expansionvector}
A_\mu = \overline A_\mu +  \sum_n B_\mu^{(n)} \psi_n.
\end{eqnarray}
To turn on an electromagnetic background field $A_\mu^{em}$ in the boundary
field theory we put ($e$ being the electromagnetic coupling constant and $Q_{em}$ the electric charge matrix)
\begin{equation} \label{Aachtergrond}
\overline A_\mu = -i e Q_{em} A_\mu^{em} = -i e  \left(
\begin{array}{cc} 2/3 & 0 \\ 0 & -1/3 \end{array} \right) A_\mu^{em} = -i e
\left( \frac{1}{6} \textbf{1}_2 + \frac{1}{2} \sigma_3 \right)  A_\mu^{em},
\end{equation}
which assigns the appropriate charge to the up- and down-quark. For the case of a constant external magnetic field along the $x_3$-direction in
the boundary field theory ($F_{12}^{em} = \partial_1 A^{em}_2 = B$), this amounts
to setting
\begin{equation}
\overline A_\mu = -i e Q_{em} x_1 B \delta_{\mu2}
=  -i x_1  \left(
\begin{array}{cc}  \frac{2}{3}eB & 0 \\ 0 &  -\frac{1}{3}eB \end{array} \right) \delta_{\mu2}
= \frac{ x_1 e B
\delta_{\mu2}}{3} \left(  -\frac{i\textbf{1}_2}{2} \right)   +  x_1 e B
\delta_{\mu2} \left(-\frac{i\sigma_3}{2}  \right),
\end{equation}
or
\begin{equation}
\overline A_2^3 = x_1 e B  \quad \mbox{   and   } \quad \overline A_2^0 =
\overline A_2^3 / 3
\end{equation}
and
\begin{eqnarray} \overline F_{12}= \partial_1 \overline A_2 =  -i \left(\begin{array}{cc} \frac{2}{3}eB  & 0 \\ 0 & -\frac{1}{3}eB \end{array} \right) = -i \left(\begin{array}{cc} \overline F_u  & 0 \\ 0 & \overline F_d \end{array} \right),   \end{eqnarray}
where in the last line we defined the up- and down-components of the background field strength, $\overline F_u$ and $\overline F_d$.

\subsection{Effect of the magnetic field on the embedding of the probe branes} \label{D}
We determine in this Section the $eB$-dependence of the embedding of the flavour D8-branes in the confining (we are working at zero temperature) D4-brane background \eqref{backgr}. On each of the D8-branes lives an induced metric
\begin{eqnarray}
ds^2_{D8} &=& g_{mn}^{D8} dx^m dx^n \quad(m,n = 0 \cdots 8) \nonumber \\
&=& \left(\frac{u}{R}\right)^{3/2} \eta_{\mu\nu}dx^\mu dx^\nu + \left( \left(\frac{R}{u}\right)^{3/2} \frac{1}{f(u)} +  \left(\frac{u}{R}\right)^{3/2} \frac{f(u)}{u'^2} \right) du^2 + \left(\frac{R}{u}\right)^{3/2} u^2 d\Omega_4^2
\end{eqnarray}
or
\begin{eqnarray}
\left( g_{00}^{D8}, g_{ii}^{D8}, g_{uu}^{D8} \right) &=& \left( -\left(\frac{u}{R}\right)^{3/2}, \left(\frac{u}{R}\right)^{3/2}, \left(\frac{u}{R}\right)^{3/2} \left[ \frac{1}{f} \left(\frac{u}{R}\right)^{-3} + \frac{f}{u'^2}\right] \right), \\
&=& \left( g_{00}, g_{ii}, G_{uu} \right) \quad \text{with $G_{uu} = g_{uu} + g_{\tau\tau}(\partial_u \tau)^2$}
 \end{eqnarray}
and a gauge field $A_\mu$, for which we assume the background gauge field ansatz
\begin{equation} \label{background}
A_\mu = \overline A_\mu =  -i e Q_{em} x_1 B \delta_{\mu 2}
\quad \mbox{(all other gauge field components zero)},
\end{equation}
modeling an external magnetic field $\vec B=B \vec e_3$ in the dual field theory.\\

We plug the gauge field ansatz into the non-Abelian DBI-action
\begin{equation}
S_{DBI} = -T_8 \int d^4x \hspace{1mm}2 \int du \int \epsilon_4 \hspace{1mm} e^{-\phi}\hspace{1mm} \text{STr}
\sqrt{-\det \left[g_{mn}^{D8} + (2\pi\alpha') iF_{mn} \right]}
\end{equation}
and solve for the embedding $u'=du/d\tau$ as a function of $eB$. To allow for the possibility that each of the two flavour D8-branes responds differently to the external magnetic field, we assume the following form of the metric in flavour space
\begin{align}
g^{D8}
&= \left( \begin{array}{cc} g^{D8}_u & 0 \\ 0 & g^{D8}_d \end{array} \right),
\end{align}
the only difference between $g^{D8}_u$ and $g^{D8}_d$ being that the $u$-coordinate appearing in $g^{D8}_u$ follows the up-brane, varying from $u_{0,u}$ to infinity, whereas the the $u$-coordinate appearing in $g^{D8}_d$ follows the down-brane, varying from $u_{0,d}$ to infinity. This will turn out to generate a different embedding $u'(eB)$ for up and down. We write
\begin{eqnarray} g^{D8} = g^{D8}  \mathbf 1', \label{metricnoncoincident}\end{eqnarray}
where we introduced the notation  
\begin{eqnarray}\mathbf 1' = \left( \begin{array}{cc} \theta(u-u_{0,u}) & 0 \\ 0 & \theta(u-u_{0,d}) \end{array} \right)\end{eqnarray}
for our ``generalized unity matrix'' that indicates that everything multiplied by the Heaviside step function $\theta(u-u_{0,u})$ (respectively $\theta(u-u_{0,d})$) will have to be integrated over $u$ varying from $u_{0,u}$ (respectively $u_{0,d})$ to infinity.\\

The determinant in the action is
\begin{align} \label{detmatrix}
\det(g^{D8}\mathbf 1' + i (2\pi\alpha')F) &= \det \left(\begin{array}{ccccc} g_{00}\mathbf 1' &0&0&0&0 \\
0 & g_{11}\mathbf 1' & i (2\pi\alpha')\overline F_{12} &0&0\\
0 & -i (2\pi\alpha')\overline F_{12} & g_{22}\mathbf 1' &0&0\\
0&0&0& g_{33}\mathbf 1' &0\\
0&0&0&0& G_{uu} \mathbf 1'
\end{array} \right)_{\text{Lorentz space}} \hspace{-1.5cm} \times \det S_4 \nonumber \\
&=  \underbrace{\det S_4 \times g_{00}g_{11}g_{22}g_{33}G_{uu}}_{\det g} \underbrace{\left(1 - (2\pi\alpha')^2 g_{11}^{-1} g_{22}^{-1} \overline F_{12}^2\right) }_{A} \mathbf 1'  \nonumber \\
&\hspace{-2cm} = \left(\begin{array}{cc}
 \det g \times A_u \times \theta(u-u_{0,u})   & 0 \\
0 &  \det g \times A_d \times \theta(u-u_{0,d}) \end{array} \right)_{\text{flavour space}},
\end{align}
where we defined
\begin{eqnarray} \label{A}
A=\left( \begin{array}{cc} A_u & 0 \\ 0 &  A_d \end{array} \right), \quad
A_{u/d} = 1 + (2\pi\alpha')^2 \overline F_{u/d}^2  \left(\frac{R}{u}\right)^{3}.
\end{eqnarray}
Since the matrix (\ref{detmatrix}) is diagonal, the square root of it is equal to the matrix of the square roots of the diagonal elements and the STr reduces to an ordinary Tr, leading to the action
\begin{equation}
S_{DBI}
= -T_8 V_{\mathbb{R}^{3+1}}  V_4 g_s^{-1} \left[ 2\int_{u_{0,u}}^\infty du \hspace{1mm} u^4
\sqrt{ \frac{1}{f} \left(\frac{u}{R}\right)^{-3} + \frac{f}{u'^2} } \sqrt{A_u} +
2\int_{u_{0,d}}^\infty du \hspace{1mm} u^4
\sqrt{ \frac{1}{f} \left(\frac{u}{R}\right)^{-3} + \frac{f}{u'^2} } \sqrt{A_d}  \right],
\end{equation}
with $V_{\mathbb{R}^{3+1}}=\int d^4x$ and $u'$ as a function of $eB$ to be determined for both the up- and down-brane.\\

Omitting all the up- and down-indices for clearness,
\begin{align}
S_{DBI} &= S_{up} + S_{down} \nonumber\\
S &=  -T_8 V_{\mathbb{R}^{3+1}} V_4 g_s^{-1}  \hspace{1mm} 2\int_{u_{0}}^\infty du \hspace{1mm} u^4 \sqrt{ \frac{1}{f} \left(\frac{u}{R}\right)^{-3} + \frac{f}{u'^2} } \sqrt{A}
\end{align}
and using the short-hand
\begin{eqnarray}
\mathcal L^\tau
= u^4  \sqrt{\frac{u'^2}{f} \left(\frac{u}{R}\right)^{-3} + f} \hspace{1mm} \sqrt{A}\,\,, \end{eqnarray}
we determine $u'$ for each flavour from the conserved ``Hamiltonian''
\begin{eqnarray} H = u' \frac{\delta \mathcal L^\tau}{\delta u'} - \mathcal L^\tau =  \frac{-u^4 f \sqrt{A}}{\sqrt{\frac{u'^2}{f} \left(\frac{u}{R}\right)^{-3} + f}}\,\,, \quad \partial_\tau H = 0. \end{eqnarray}
Expressing that this $H$ is conserved and assuming a $\cup$-shaped embedding $u' = 0$ at $u=u_{0}$ (with $A(u_0)$ and $f(u_0)$ denoted as $A_0$ and $f_0$):
\begin{eqnarray}
\frac{-u^4 f \sqrt{A}}{\sqrt{\frac{u'^2}{f} \left(\frac{u}{R}\right)^{-3} + f}}
=
\frac{-u_0^4 f_0 \sqrt{A_0}}{\sqrt{f_0}}
\end{eqnarray}
we find
\begin{equation}
u'^2 = \left(\frac{u}{R}\right)^{3} f^2 \frac{u^8 f A - u_0^8 f_0 A_0}{u_0^8 f_0 A_0},
\end{equation}
reducing to the known $\cup$-shaped embedding for $eB\rightarrow 0$ whereby $A\rightarrow 1$.

\subsubsection[Antipodal embedding]{Antipodal embedding ($u_0=u_K$): no dependence on $eB$}

In the case $u_0=u_K$, we have $f_0=0$ so
\begin{eqnarray}
(\partial_u \tau)^2 = \left(\frac{R}{u}\right)^{3} \frac{1}{f^2} \frac{u_0^8 f_0 A_0}{u^8 f A - u_0^8 f_0 A_0} = 0
\end{eqnarray}
and the embedding function is constant,
\begin{eqnarray}
\tau(u) = \overline \tau \sim \mathbf 1,
\end{eqnarray}
($\overline \tau = 0$ for the D8-branes and $\overline \tau=\pi/M_K \mathbf 1$ for the $\overline{\text{D8}}$-branes, see the l.h.s.~of Figure  \ref{SS}), independent of the value of the magnetic field. In this case, there is thus no response of the chiral symmetry breaking to the magnetic field, a somewhat unphysical feature of the extremal Sakai-Sugimoto embedding, which is a direct consequence of the absence of a constituent quark mass in this setting.

\subsubsection[Non-antipodal embedding]{Non-antipodal embedding ($u_0>u_K$): magnetic catalysis of chiral symmetry breaking}

\begin{figure}[h!]
  \centering
  \scalebox{0.8}{
  \includegraphics{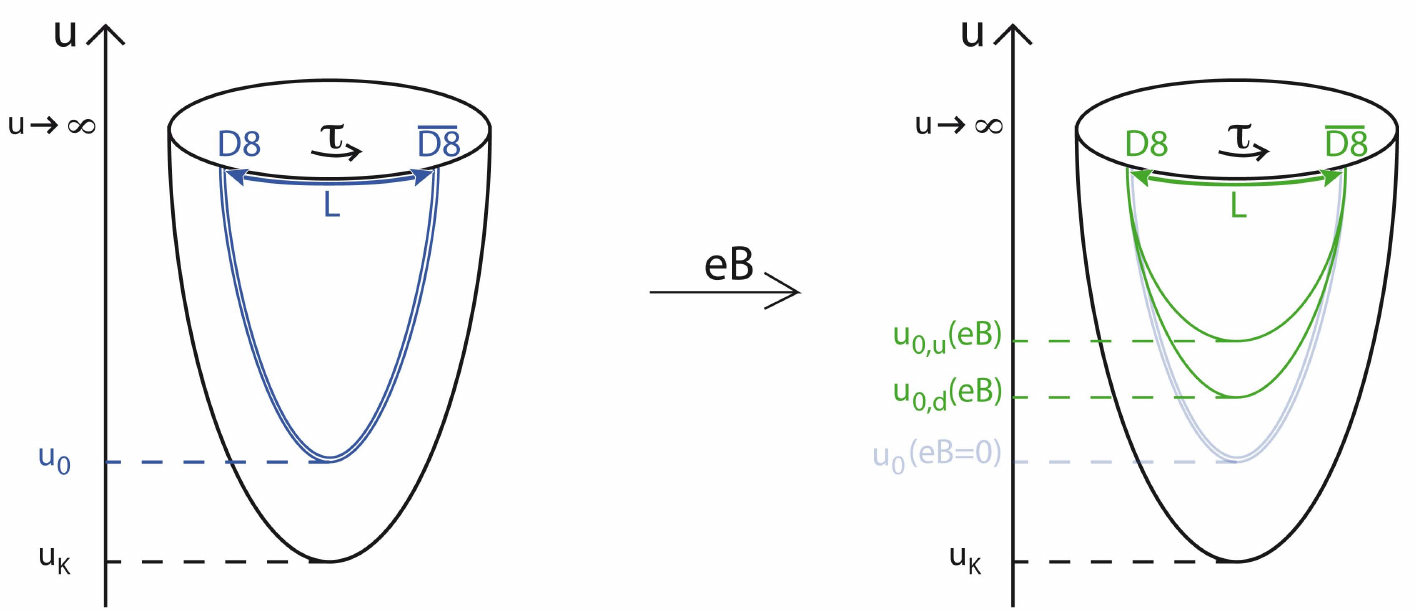}}
  \caption{The change in embedding of the flavour branes caused by the magnetic field $eB$ models the chiral magnetic catalysis effect. The up-brane reacts the strongest to $eB$, corresponding to a stronger chiral magnetic catalysis for the up-quarks than for the down-quarks.}\label{changedembedding}
\end{figure}

In the case $u_0 > u_K$ the embedding function for the flavour branes in the background is given by
\begin{eqnarray}
\tau(u) = \overline \tau =\left( \begin{array}{cc} \overline \tau_u & 0 \\ 0 &  \overline \tau_d \end{array} \right)
\end{eqnarray}
with
\begin{equation} \label{tauembedding}
\partial_u \overline \tau = \sqrt{ \left(\frac{R}{u}\right)^{3} \frac{1}{f^2} \frac{u_0^8 f_0 A_0}{u^8 f A - u_0^8 f_0 A_0}}  \times \theta(u - u_0), 
\end{equation}
with $u_0$ and $A$ taking their up or down values in $\p_u \overline \tau_u$ or $\p_u \overline \tau_d$ respectively. The up-brane and down-brane are thus no longer coincident in the presence of a magnetic field, as sketched in Figure \ref{changedembedding}.   \\ 

The $eB$-dependent induced metric on the up- and down-brane is given by
\begin{eqnarray} \left( g_{00}^{D8}, g_{ii}^{D8}, g_{uu}^{D8} \right) = \left( -\left(\frac{u}{R}\right)^{3/2}, \left(\frac{u}{R}\right)^{3/2}, \left(\frac{R}{u}\right)^{3/2} \gamma_B(u) \right)  \end{eqnarray}
and the action by
\begin{align}
S_{DBI} &= S_{up} + S_{down}, \nonumber\\
S &=  -T_8 V_{\mathbb{R}^{3+1}} V_4 g_s^{-1} \hspace{1mm} 2 \int_{u_0}^\infty du \hspace{1mm}  R^{3/2} u^{5/2} \sqrt{A} \sqrt{\gamma_B}  \label{actionmagneticfield}
\end{align}
with
\begin{eqnarray} \label{gammaB}
\gamma_B(u) = \frac{u^8 A}{u^8 f A - u_0^8 f_0 A_0}.
\end{eqnarray}
We see that the non-Abelian DBI-action for the two D8-branes in the presence of a background magnetic field reduces to the sum of two Abelian actions. This represents the explicit breaking of the global chiral symmetry,
\begin{eqnarray}
U(2)_L \times U(2)_R \stackrel{eB}{\rightarrow} (U(1)_L \times U(1)_R)^u \times (U(1)_L \times U(1)_R)^d,
\end{eqnarray}
caused by the up- and down-quarks' different coupling to the magnetic field. \\

The asymptotic separation $L$ between D8- and $\overline{\mbox{D8}}$-branes
as a function of the magnetic field is
\begin{align}
L     &= 2 \int_{u_0}^\infty du  \left(\frac{R}{u}\right)^{3/2} f^{-1} \sqrt{\frac{u_0^8 f_0  A_0}{u^8 f A - u_0^8 f_0  A_0}} \nonumber \\
   &= \frac{2}{3} \frac{R^{3/2}}{\sqrt {u_0}} \sqrt{f_0 A_0} \int_0^1 d\zeta \frac{f^{-1} \zeta^{1/2}}{\sqrt{fA - f_0 A_0 \zeta^{8/3}}} \label{LconfifvB}
\end{align}
where we changed the integration variable to $\zeta = (u/u_0)^{-3}$ \cite{Johnson:2008vna},
with $y_K = u_K/u_0$, $y=u/u_0$ and $f = 1 - y_K^3 \zeta$.\\

In Section \ref{B} the value of the geometric parameter $L$ in zero magnetic field was determined at $L(eB=0) = 1.547$ GeV$^{-1}$.
We keep $L = L(eB=0)$ fixed
while varying $eB$ to determine $u_{0,u/d}(eB)$ and consequently, via
\begin{equation}\label{masscont}
m_q(M_K,u_0,\kappa) =8\pi^2 M_K^2 \kappa \int_{1/M_K}^{u_0} du \frac{1}{\sqrt{1 - \frac{1}{(M_Ku)^3}}},
\end{equation}
the constituent quark masses $m_{u}(eB)$ and $m_d(eB)$ of up- and down-quarks. 
\\

\begin{figure}[h!]
  \hfill
    \begin{center}
      \scalebox{0.7}{
  \includegraphics{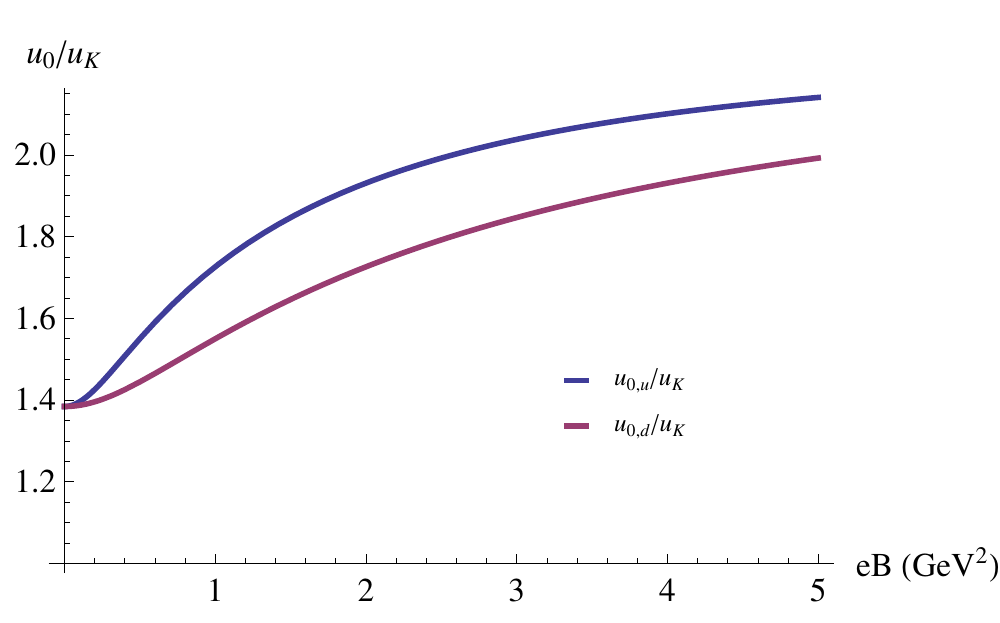}}
     \end{center}
      \caption{$\frac{u_0}{u_K}$ as a function of the magnetic field for the D8-brane corresponding to the up-quark, and the one corresponding to the down-quark.}
	\label{u0}
  \hfill
\end{figure}

\begin{figure}[h!]
  \hfill
  \begin{minipage}[t]{.45\textwidth}
    \begin{center}
      \scalebox{0.7}{
  \includegraphics{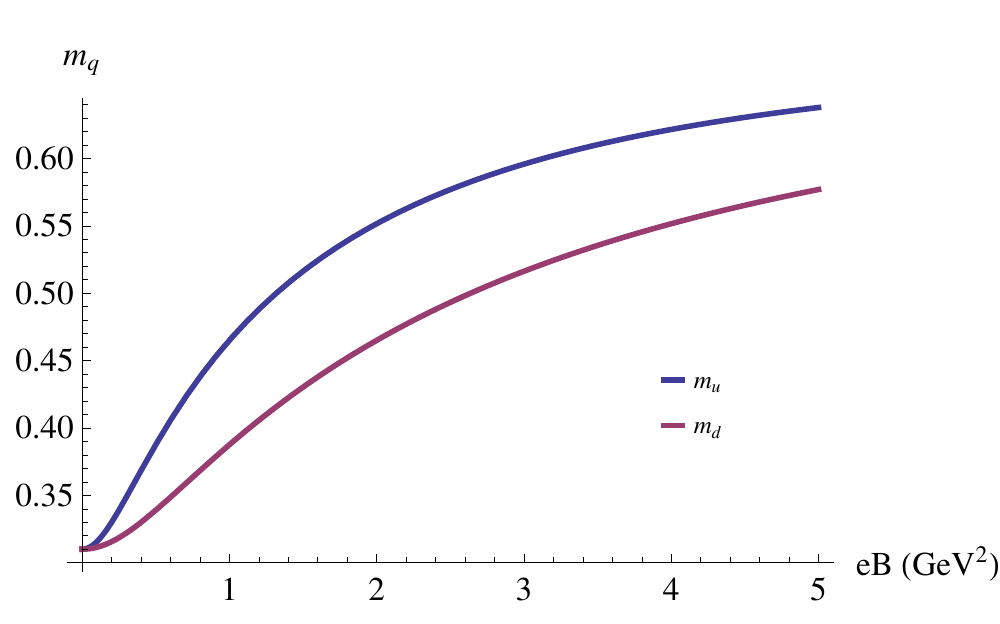}}
    \end{center}
  \end{minipage}
  \hfill
  \begin{minipage}[t]{.45\textwidth}
    \begin{center}
      \scalebox{0.7}{
  \includegraphics{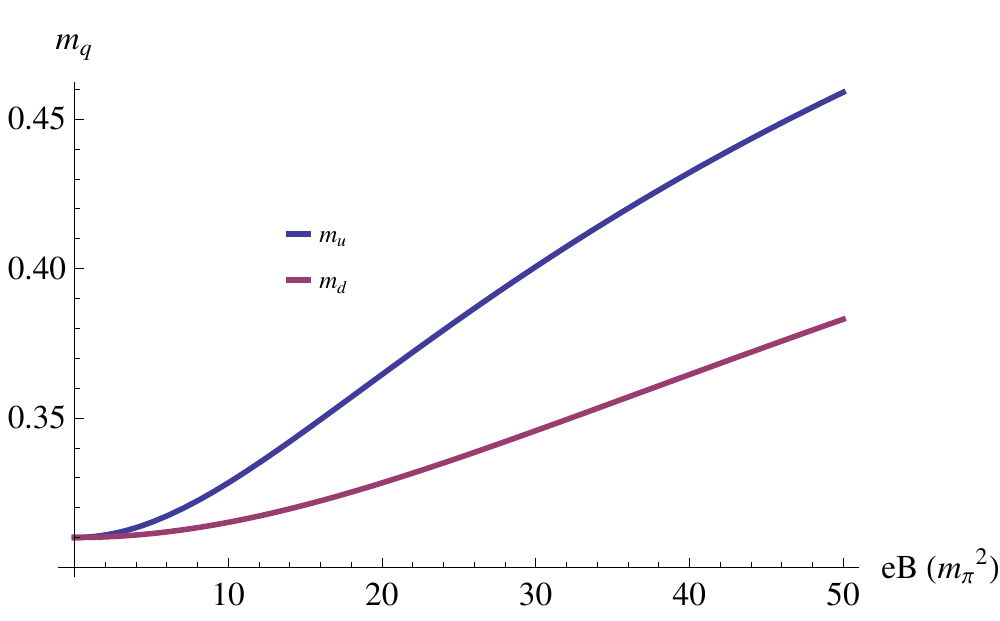}}
    \end{center}
  \end{minipage}
      \caption{The constituent masses of the up-quark and the down-quark as a function of the magnetic field (in units GeV$^{2}$ and $m_\pi^2 = (0.138$ GeV)$^2$).}
	\label{mqfig}
  \hfill
\end{figure}

In Figures \ref{u0} and \ref{mqfig} the numerically obtained dependence on $eB$ of $u_{0,u/d}$, $m_{u}$ and $m_d$ are depicted. As $u_0$ rises with $eB$, the probe branes in the presence of the external magnetic field get more and more bent
towards each other, driving them further and further away from the chirally invariant situation of straight branes, see Figure \ref{changedembedding}. This feature corresponds to a holographic modeling of the magnetic catalysis of chiral symmetry breaking \cite{Miransky:2002rp}. As already mentioned, this ``chiral magnetic catalysis effect'' was already discussed for the Sakai-Sugimoto model in \cite{Johnson:2008vna}, albeit for a single flavour and without matching the free parameters onto QCD values. The constituent quark masses $m_q(eB)$, which are related to the quantity $u_0(eB) - u_K$, accordingly increase (see Figure \ref{mqfig}), leading us to expect that taking this chiral magnetic catalysis into account will translate into the $\rho$ meson mass also growing with $eB$, at least when ignoring the lowest Landau level shift (see next Section). \\

\begin{figure}[h!]
  \hfill
  \begin{minipage}[t]{.45\textwidth}
    \begin{center}
      \scalebox{0.8}{
  \includegraphics{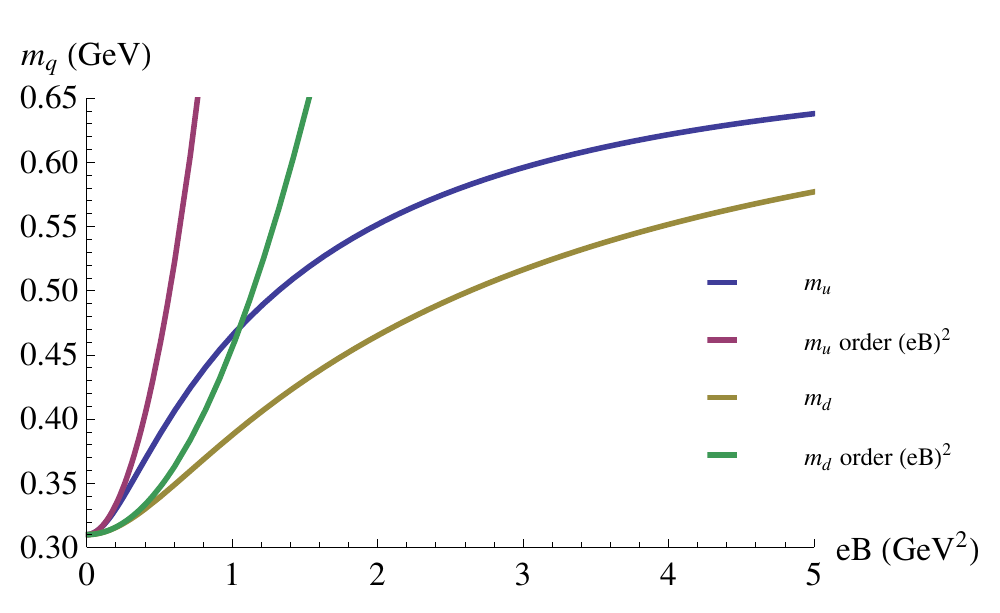}}
    \end{center}
  \end{minipage}
  \hfill
  \begin{minipage}[t]{.45\textwidth}
    \begin{center}
      \scalebox{0.8}{
  \includegraphics{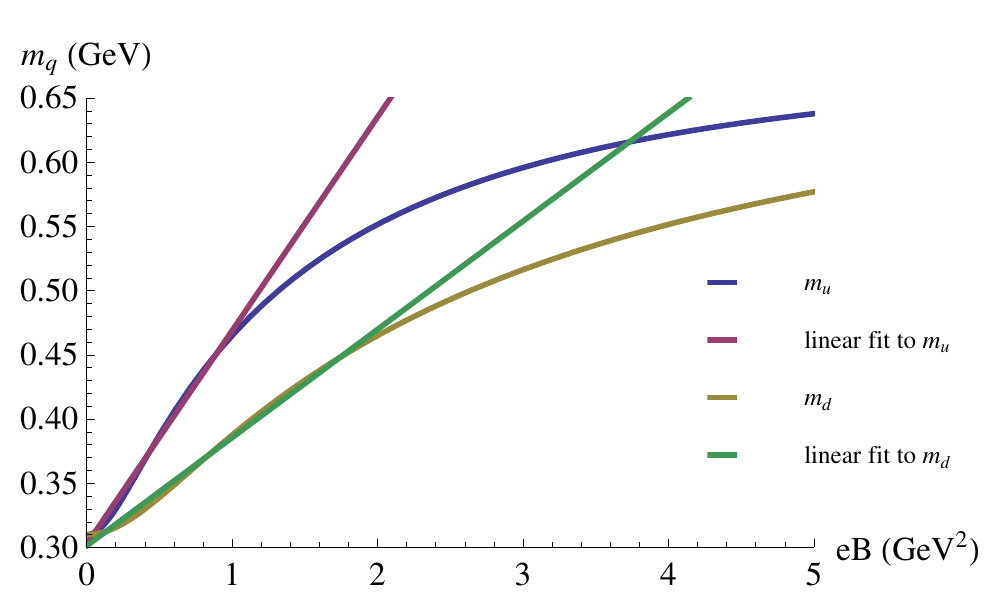}}
    \end{center}
  \end{minipage}
      \caption{Left: the constituent up-quark and down-quark masses and their quadratic approximation \eqref{kwad}. Right: a linear fit \eqref{fit} to the same quantities.}
	\label{mqkwadr}
  \hfill
\end{figure}

For small magnetic fields, we can compute the lowest order correction to $m_q(eB=0)$ analytically, confirming the typical holographic $(eB)^2$ dependence \cite{Zayakin:2008cy}.
To this end we approximate $L$ by $L = \ell_0(u) + (eB)^2 \, \ell_1(u)$. The condition $\ell_0(u) + (eB)^2 \, \ell_1(u)  = L(eB=0)$ then has a solution of the form $u = u_0 + (eB)^2 \, u_1$ with $u_0 =
1.38 \, u_K = 1.92 \text{ GeV}^{-1}$ and $u_1 = -\ell_1(u_0) \left(\frac{d\ell_0}{du}(u_0)\right)^{-1}$. The corresponding constituent quark masses for small magnetic fields are
\begin{align}
 m_q(eB) &= 0.310 \mbox{  GeV} + 8\pi^2 M_K^2 \kappa \int_{u_0}^{u_0 + (eB)^2 \, u_1} du \frac{1}{\sqrt{1 - \frac{1}{(M_Ku)^3}}} \nonumber\\
&= 0.310 \mbox{  GeV} + (8\pi^2 M_K^2 \kappa) (eB)^2 \, u_1 \frac{1}{\sqrt{1 - \frac{1}{(M_K u_0)^3}}}.
\end{align}
We find
\begin{equation}\label{kwad}
m_u(eB) =  0.310 \mbox{  GeV} + 0.582 ~ (eB)^2 + \mathcal{O}(eB)^3 \quad \mbox{ and } \quad m_d(eB) =  0.310 \mbox{  GeV} + 0.145 ~ (eB)^2+ \mathcal{O}(eB)^3,
\end{equation}
depicted in Figure \ref{mqkwadr}. This quadratic dependence at small magnetic field is also encountered in other effective descriptions of the constituent quark mass, as in the $\text{PLSM}_q$ model of \cite{Mizher:2011wd,Mizher:2011wdb}, and although not explicitly mentioned, also an instanton based computation seems to give a quadratic-like power \cite{Nam:2011vn}. Also the numerical data of \cite{D'Elia:2011zu} for the up- and down-quark chiral condensates are in accordance with a quadratic behaviour at small $eB$. We must however mention that the latter lattice computations were done at nonvanishing current quark mass in an unquenched setting. Chiral perturbation theory predicts a linear behaviour \cite{Shushpanov:1997sf} (see also the comments in \cite{Zayakin:2008cy}). Quenched lattice simulations of \cite{Buividovich:2008wf} confirmed this, although we notice that the small $eB$-behaviour does not seem to be precisely caught by the proposed linear fit. In fact, we are able to fit our result quite well with a linear fit if $eB$ is not too large, see Figure \ref{mqkwadr}. We used
\begin{equation}\label{fit}
m_u(eB)^{linear} \approx 0.303~\text{GeV}+0.166 ~ eB \quad \mbox{ and } \quad m_d(eB)^{linear} \approx  0.301~\text{GeV} + 0.084 ~ eB.
\end{equation}
It is also instructive to see what happens at large magnetic field. As already pointed out in \cite{Johnson:2008vna}, we observe a saturation in Figure \ref{mqfig}. Such a saturation was also seen for the first time using lattice simulations in \cite{D'Elia:2011zu} for the up- and down-quark chiral condensates, carefully taken into account some subtleties related to an unphysical periodicity in the results, which is a typical lattice artefact. The results of \cite{Nam:2011vn,D'Elia:2011zu} anyhow confirm the different response of the constituent up- and down-quark masses or chiral condensates to the magnetic field, a feature which we also reproduced here for the first time in the holographic Sakai-Sugimoto setting.  It thus appears that our holographic results reproduce quite well the phenomenology of independent quenched QCD calculations. The unquenched top-of-the-bill simulations of \cite{Bali:2012zg} show a similar behaviour for the chiral condensate, at least at vanishing temperature. Similarly shaped curves for the chiral condensate extrapolated to the chiral limit can be found in \cite{Ilgenfritz:2012fw}, which would be most relevant for comparison with our analysis, nevertheless keeping in mind that the lattice study \cite{Ilgenfritz:2012fw} is for 2 rather than 3 colours.

\section{Exploring the $\rho$ meson mass}
In this section we will determine the mass equation for the 5-dimensional gauge field fluctuation describing the four-dimensional $\rho$ meson, again for both the antipodal and the non-antipodal embedding.
The strategy is to plug the gauge field ansatz
\begin{equation}
\left\{\begin{array}{ll} A_r = \overline A_r + \tilde A_r  \quad (r=\mu,u) \\ \tau = \overline \tau + \tilde \tau  \end{array} \right.
\end{equation}
with (see (\ref{background}) and (\ref{tauembedding}))
\begin{equation} 
\left\{\begin{array}{ll}  \overline A_\mu = -i e Q_{em} x_1 B \delta_{\mu2}
\\ \partial_u\overline \tau =  \sqrt{\left(\frac{R}{u}\right)^{3} \frac{1}{f^2} \frac{u_0^8 f_0 A_0}{u^8 f A - u_0^8 f_0 A_0}} \times \theta(u-u_0)
\end{array} \right.,
\end{equation}
into the DBI-action describing the dynamics of the flavour branes. We then expand the action to order $(2\pi\alpha')^2 \sim \frac{1}{\lambda^2}$ and to second order in the fluctuations, and finally integrate out the $u$-dependence to obtain an effective four-dimensional theory for the $\rho$ meson in the presence of the background magnetic field. Further we will use the notation $\overline \tau = \overline \tau^a \sigma^a$, $\tilde \tau = \tilde \tau^a \sigma^a$, $\overline A_\mu = \overline A_\mu^a t^a$ and $\tilde A_r = \tilde A_r^a t^a$, with all components field$^a$ real.\\

We have to evaluate (for notational brevity we temporarily absorb the factor $(2\pi\alpha')$ into the field strength)
\begin{equation} \label{tensionabsorbed}
2 \int du \hspace{2mm} \text{STr} \sqrt{-\det (a_{rs})} = 2 \int du \hspace{2mm}  \text{STr} \sqrt{-\det (g_{rs}^{D8} + i F_{rs})},
\end{equation}
with
\begin{align}
 g_{rs}^{D8} &= g_{rs} + g_{\tau\tau} D_r \tau D_s \tau, \qquad \mbox{with $D_r \cdot = \partial_r + [\overline A_r,\cdot]$}
 \end{align}
and
\begin{align}
F_{rs} &= \partial_r A_s - \partial_s A_r + [A_r, A_s].
\end{align}
If we denote the argument $a$ of the determinant (which runs over the Lorentz-indices) as
\[ a = \overline a + a^{(1)} + a^{(2)} + \cdots \]
with  $a^{(n)}$ being $n$-th order in the fluctuations $\tilde A$,
the determinant can be expanded to second order in the fluctuations as follows
\begin{align}
\sqrt{-\det a}|_{\tilde A^2} &= \sqrt{-\det \overline a}  \left\{ 1+\frac{1}{2} \text{tr} (\overline a^{-1} a^{(1)}) + \frac{1}{8} \left( \text{tr}(\overline a^{-1} a^{(1)}) \right)^2 - \frac{1}{4} \text{tr}\left( (\overline a^{-1} a^{(1)})^2 \right) + \frac{1}{2} \text{tr} (\overline a^{-1} a^{(2)}) \right\}.  \label{expansion}
\end{align}
We denote the trace in Lorentz-space with a small $\text{tr}$, and the trace in flavour space with a capital (S)Tr.\\

Splitting each component of $a$ in its symmetric and antisymmetric parts
\begin{equation} \label{a}
\left\{ \begin{array}{ll} \overline a^{-1} = \mathcal G + \mathcal B \\
 					    a^{(1)} = \overline a^{(1)} + \delta_1 F \\
					    a^{(2)} = \overline a^{(2)} + \delta_2 F \\
\end{array} \right.
\end{equation}
the expansion of the determinant (\ref{expansion}) to second order in the fluctuations can be written as
\begin{align}
\sqrt{-\det a} |_{\tilde A^2}
&= \sqrt{-\det \overline a} + \sqrt{-\det \overline a} \times \nonumber\\
 & \hspace{-1cm}\left\{ \frac{1}{2}  \text{tr} ( \mathcal G \overline a^{(1)})  + \frac{1}{8}  \left(\text{tr} ( \mathcal G \overline a^{(1)})\right)^2  - \frac{1}{4}  \text{tr} ( \mathcal G \overline a^{(1)}\mathcal G \overline a^{(1)} + \mathcal B \overline a^{(1)}\mathcal B \overline a^{(1)}) + \frac{1}{2}  \text{tr} ( \mathcal G \overline a^{(2)}) \right.  \nonumber\\ 
& \hspace{-1cm} +\frac{1}{2}  \text{tr} ( \mathcal B \delta_1 F)  + \frac{1}{8}  \left(\text{tr} (\mathcal B \delta_1 F)\right)^2  - \frac{1}{4}  \text{tr} (\mathcal G \delta_1 F\mathcal G \delta_1 F + \mathcal B \delta_1 F\mathcal B \delta_1 F) + \frac{1}{2}  \text{tr} (\mathcal B \delta_2 F)   \nonumber \\ 
 &\hspace{-1cm} \left. +\frac{1}{4} \text{tr} ( \mathcal G \overline a^{(1)})\text{tr} ( \mathcal B \delta_1 F) - \frac{1}{2} \text{tr}(\mathcal G \overline a^{(1)} \mathcal B \delta_1 F) - \frac{1}{2} \text{tr}(
 \mathcal G \delta_1 F \mathcal B \overline a^{(1)}) \right\}. \label{derdelijnscalarplusvector}
\end{align}

For our field ansatz we have
\begin{align}
\overline a_{rs} &= g_{rs} + g_{\tau\tau} \partial_r \overline \tau \partial_s \overline \tau +  i \overline F_{rs}, \\
\overline a^{(1)}_{rs} &= g_{\tau\tau} \left(\partial_r \overline\tau \left([\tilde A_s,\overline\tau] + D_s \tilde \tau \right) + \left([\tilde A_r,\overline\tau] + D_r \tilde \tau\right) \partial_s \overline\tau \right), \\
\delta_1 F_{rs} &= i (D_r \tilde A_s - D_s \tilde A_r) \stackrel{notation}{=} i \tilde F_{rs} \label{notF1}\\
\overline a^{(2)}_{rs} &= g_{\tau\tau}\left([\tilde A_r,\overline\tau]+D_r \tilde \tau\right) \left([\tilde A_s,\overline\tau]+D_s \tilde \tau\right) + g_{\tau\tau} \left([\tilde A_r,\tilde \tau] \partial_s \overline \tau+  \partial_r \overline \tau [\tilde A_r,\tilde \tau] \right), \\
\delta_2 F_{rs} &= i [\tilde A_r, \tilde A_s]. 
\end{align}
The symmetric part $\mathcal G$ of $\overline a^{-1}$ is diagonal,
\begin{equation}
\mathcal G = \left( \begin{array}{ccccc}g_{00}^{-1} & & & & \\ & g_{11}^{-1} A^{-1} & & & \\ & & g_{22}^{-1} A^{-1} & & \\ & & & g_{33}^{-1} & \\ & & & & G_{uu}^{-1} \end{array} \right), \quad \mbox{ with } G_{uu} = g_{uu}+g_{\tau\tau} (\partial_u \tau)^2
\end{equation}
and the antisymmetric part $\mathcal B$ has non-zero components
\begin{equation}
\mathcal B_{12}=-\mathcal B_{21} = i \overline F_{12} g_{11}^{-1} g_{22}^{-1} A^{-1}.
\end{equation}
The DBI-Lagrangian to second order in the fluctuations thus becomes
\begin{align}
\text{STr} & \hspace{1mm} e^{-\phi} \sqrt{-\det a} |_{\tilde A^2,\tilde \tau^2,\tilde A \tilde \tau}
=  \text{Tr} \hspace{1mm} e^{-\phi} \sqrt{-\det \overline a} + \text{STr} \hspace{1mm} e^{-\phi} g_{11}^2 \sqrt{G_{uu}} g_{S_4}^2 \sqrt{A} \times \nonumber\\
 & \hspace{-2cm}\left\{
 \frac{1}{2} \left( [\tilde A_u,\overline \tau] +D_u\tilde\tau \right)^2 G_{uu}^{-2}
 + [\tilde A_u,\tilde \tau] (G_{uu}^{-1}g_{\tau\tau}\partial_u \overline \tau) \right.
+ \frac{1}{2} \left( [\tilde A_\mu,\overline \tau] +D_\mu \tilde\tau \right)^2 g_{\mu\mu}^{-1} A^{-1}|_{\mu=1,2} G_{uu}^{-1}
 \nonumber \\
& \hspace{-2cm} + \overline F_{12} g_{11}^{-2} A^{-1} \tilde F_{21}
- \overline F_{12} g_{11}^{-2} A^{-1} [\tilde A_1,\tilde A_2]
- \frac{1}{4} g_{\mu\mu}^{-1} A^{-1}|_{\mu=1,2} g_{\nu\nu}^{-1} A^{-1}|_{\nu=1,2} \tilde F_{\mu\nu}^2 - \frac{1}{2} g_{\mu\mu}^{-1} A^{-1}|_{\mu=1,2} G_{uu}^{-1} \tilde F_{\mu u}^2 \nonumber \\
& \left. \hspace{-2cm}
- G_{uu}^{-1} g_{\tau\tau} \partial_u \overline \tau \overline F_{12}g_{11}^{-2} A^{-1} \left( \left( [\tilde A_u,\overline \tau]+ D_u \tilde\tau \right) \tilde  F_{12} + \left( [\tilde A_1,\overline \tau]+ D_1 \tilde\tau \right)  \tilde F_{2u} - \left( [\tilde A_2,\overline \tau]+ D_2 \tilde\tau \right) \tilde F_{1u} \right)
\right\} \label{}
\end{align}
where the notation for the factors $g_{\mu\mu}^{-1} A^{-1}|_{\mu=1,2}$ coming from $\mathcal G$ means that $g_{\mu\mu}^{-1}$ is accompanied with a factor $A^{-1}$ only for $\mu=1,2$.

Approximating the action to order $T^{-2} = (2\pi\alpha')^2 \sim 1/\lambda^2$ while keeping in mind that every $F$ carries a factor $T^{-1}$ and that (see (\ref{detmatrix})) 
\begin{align}
A  &= 1 - g_{11}^{-2} T^{-2} \overline F_{12}^2,
\end{align}
we have
\begin{align}
\text{STr} & \hspace{1mm} e^{-\phi} \sqrt{-\det a} |_{\tilde A^2,\tilde \tau^2,\tilde A \tilde \tau,T^{-2}}
= \text{Tr} \hspace{1mm} e^{-\phi} \sqrt{-\det \overline a} + \text{STr}  \hspace{1mm} e^{-\phi} g_{11}^2 \sqrt{G_{uu}} g_{S_4}^2 T^{-2}  \times \nonumber\\
 & \left\{
\left(T^2 -\frac{1}{2} g_{11}^{-2} \overline F_{12}^2 \right) \left\{\frac{1}{2} \left( [\tilde A_u,\overline \tau] +D_u\tilde\tau \right)^2 G_{uu}^{-2}
 + [\tilde A_u,\tilde \tau] (G_{uu}^{-1}g_{\tau\tau}\partial_u \overline \tau)
 \right\} \right. \nonumber\\
&
+ \left(T^2 - \frac{1}{2} g_{11}^{-2} \overline F_{12}^2 \right) \frac{1}{2} \left( [\tilde A_\mu,\overline \tau] +D_\mu \tilde\tau \right)^2 g_{\mu\mu}^{-1} G_{uu}^{-1} + \sum_{\mu=1}^2 \left(g_{11}^{-2} \overline F_{12}^2 \right) \frac{1}{2} \left( [\tilde A_\mu,\overline \tau] +D_\mu \tilde\tau \right)^2 g_{\mu\mu}^{-1} G_{uu}^{-1} \nonumber \\
& 
+ \overline F_{12} g_{11}^{-2} \tilde F_{21}
- \overline F_{12} g_{11}^{-2} [\tilde A_1,\tilde A_2]
- \frac{1}{4} g_{\mu\mu}^{-1} g_{\nu\nu}^{-1}\tilde F_{\mu\nu}^2 - \frac{1}{2} g_{\mu\mu}^{-1} G_{uu}^{-1} \tilde F_{\mu u}^2 \nonumber \\
& \left. 
- G_{uu}^{-1} g_{\tau\tau} \partial_u \overline \tau \overline F_{12}g_{11}^{-2} \left( \left( [\tilde A_u,\overline \tau]+ D_u \tilde\tau \right) \tilde F_{12} + \left( [\tilde A_1,\overline \tau]+ D_1 \tilde\tau \right) \tilde F_{2u} - \left( [\tilde A_2,\overline \tau]+ D_2 \tilde\tau \right) \tilde F_{1u} \right)
\right\}. \label{detscalarplusvectorT}
\end{align}

\subsection[Antipodal embedding]{Antipodal embedding ($u_0=u_K$): Landau levels $\Rightarrow eB_c = m_\rho^2$} \label{previous}

In the case $u_0=u_K$ we have
\begin{align}
 \partial_u \overline \tau &= 0 \Rightarrow  G_{uu} = g_{uu} \\
 \overline \tau &\sim \mathbf{1} \Rightarrow [A_r,\overline \tau] = 0,
\end{align}
simplifying (\ref{detscalarplusvectorT}) drastically to a part in the scalar fluctuations and a part quadratic in the gauge fluctuations, given by
\begin{align}
\text{STr} & \hspace{1mm} e^{-\phi} \sqrt{-\det a} |_{\tilde A^2,T^{-2}}
= \text{Tr}  \hspace{1mm} e^{-\phi} g_{11}^2 \sqrt{g_{uu}} g_{S_4}^2 T^{-2}  \times  \nonumber\\ &
\left\{
\overline F_{12} g_{11}^{-2} \tilde F_{21}
- \overline F_{12} g_{11}^{-2} [\tilde A_1,\tilde A_2]
- \frac{1}{4} g_{\mu\mu}^{-1} g_{\nu\nu}^{-1} \tilde F_{\mu\nu}^2 - \frac{1}{2} g_{\mu\mu}^{-1} g_{uu}^{-1} \tilde F_{\mu u}^2 \right\},
\end{align}
where we also chose the gauge $A_u=0$.
The DBI-action for the $A_\mu^{a}$ ($a=1,2$) components can then be written as\footnote{Note that from here on contraction over Minkowski indices is assumed implicitly in the notation of squares, e.g. $(\tilde F_{\mu\nu}^a)^2 =  F_{\mu\nu}^a  F_{\mu\nu}^a \eta^{\mu\mu} \eta^{\nu\nu}$.} 
\begin{align}
S_{DBI} &= \int d^4 x \int du \sum_{a=1}^2 \left\{ -\frac{1}{4} f_1 (\tilde F_{\mu\nu}^a)^2 - \frac{1}{2} f_2 (\tilde F_{\mu u}^a)^2 - \frac{1}{2} f_3 \sum_{\mu,\nu=1}^2 \overline F_{\mu\nu}^3 \epsilon_{3ab} \tilde A^{\mu a} \tilde A^{\nu b} \right\} \nonumber\\
&=  \int d^4x \int du \sum_{a=1}^2 \left\{ -\frac{1}{4} f_1(\mathcal F_{\mu\nu}^a)^2 \psi^2 - \frac{1}{2} f_2(\rho_\mu^a)^2 (\partial_u \psi)^2 - \frac{1}{2} f_3 \sum_{\mu,\nu=1}^2 \overline F_{\mu\nu}^3 \epsilon_{3ab} \rho^{\mu a} \rho^{\nu b} \psi^2 \right\} + \text{pion action}
\end{align}
with $f_i$ ($i=1..3$) the following functions of $u$:
\begin{eqnarray}
f_1(u) = f_3(u)  &=& T_8 V_4 (2\pi\alpha')^2 e^{-\phi} g^2_{S_4} \sqrt{g_{uu}} \\
f_2(u) &=& T_8 V_4 (2\pi\alpha')^2 e^{-\phi} g^2_{S_4} g_{11} g_{uu}^{-1/2},
\end{eqnarray}
and where we retained only the first meson of the vector meson tower in the fluctuation expansion (\ref{expansionvectorpion}), $\tilde A_\mu^a = \rho_\mu^a(x^\mu) \psi(u) + \text{pions}$ ($a=1,2$), because it is the most likely to condense first, being the lightest spin 1 particle, and $\mathcal F_{\mu\nu}^a = D_\mu \rho_\nu^a - D_\nu \rho_\mu^a$.
There are no coupling terms between $\rho$ mesons and pions at second order in the fluctuations in the DBI-action, which can be traced back to the different parity of $\psi_0(z) \equiv \psi_0(u(z))$ and $\psi(z) \equiv \psi(u(z))$ (with $u(z) = u_K^3 + u_K z^2$), $\psi_0(z)$ being odd and $\psi(z)$ even.
In order to obtain a canonical kinetic term and mass term for the $\rho$ mesons in the effective four-dimensional action, we impose that the $\psi(u)$ fulfill the standard conditions
\begin{align}
\int_{u_K}^\infty du \hspace{1mm} f_1 \psi^2 &= 1\,, \\
\int_{u_K}^\infty du \hspace{1mm} f_2 (\partial_u \psi)^2 &= m_\rho^2.
\end{align}
These conditions combine to the eigenvalue equation (\ref{finiteTeigwvgl}) with $\gamma = f^{-1}$ for $u_0=u_K$. Per construction the lowest eigenvalue of this equation is $m_\rho^2 = 0.776^2$ GeV$^2$. The corresponding eigenfunction $\psi$ fulfilling the boundary conditions $\psi'(z=0) = 0$ and $\psi(z\rightarrow\pm\infty)=0$ can be used to evaluate the last integral over $u$ in the above action, which determines the non-minimal magnetic moment coupling $k$ of the $\rho$ mesons to the background magnetic field,
\begin{equation}
\int_{u_K}^\infty du \hspace{1mm} f_3 \psi^2 = k,
\end{equation}
related to the magnetic moment $\mu$ as $\mu = (1+k) e /(2m)$ so to the gyromagnetic ratio $g$ as $g=1+k$ \cite{Obukhov:1984xb}. Because in this simple embedding we have $f_1 = f_3$, we immediately see from the normalization condition that $k=1$ and thus $g=2$, describing a non-minimal coupling of the $\rho$ mesons to the background magnetic field.\\

The effective four-dimensional action thus takes the form of the standard four-dimensional action used to describe the coupling of charged vector mesons to an external magnetic field (i.e. the Proca action \cite{Obukhov:1984xb,Tsai:1972iq} or DSGS action for self-consistent $\rho$ meson quantum electrodynamics to second order in the fields \cite{Djukanovic:2005ag}):
\begin{equation}
S_{eff} = \int d^4x \sum_{a=1}^2  \left\{ -\frac{1}{4}(\mathcal F_{\mu\nu}^a)^2 - \frac{1}{2} m_\rho^2 (\rho_\mu^a)^2  - \frac{1}{2} \sum_{\mu,\nu=1}^2 \overline F_{\mu\nu}^3 \epsilon_{3ab} \rho^{\mu a} \rho^{\nu b} \right\}.
\end{equation}
This means that the Sakai-Sugimoto model with $u_0=u_K$ automatically describes Landau levels for $\rho$ mesons moving in an external magnetic field.
Let us quickly repeat how to derive this from the effective action, following for example \cite{Obukhov:1984xb} (up to conventions).\\

The equations of motion for $\rho^{\nu a}$,
\begin{equation}
D^\mu \mathcal F_{\mu\nu}^a - \epsilon_{a3b} \overline F_{\mu\nu}^3 \rho^{\mu b} - m^2 \rho_\nu^a = 0,
\end{equation}
combine to
\begin{equation}
\mathbf{D}^\mu(\mathbf D_\mu \rho_\nu - \mathbf D_\nu \rho_\mu) - i \overline F_{\mu\nu}^3 \rho^\mu - m^2 \rho_\nu = 0
\end{equation}
with $\mathbf D_\mu = \partial_\mu + i \overline A_\mu^3 = \partial_\mu + i e \overline A_\mu^{em}$ for the charged combination $\rho_\mu^- = (\rho_\mu^1 + i \rho_\mu^2)/\sqrt 2$, and the complex conjugate of this equation for the other charged combination $\rho_\mu^+ = (\rho_\mu^1 - i \rho_\mu^2)/\sqrt 2$. Acting with $\mathbf D^\nu$ on this equation of motion, and using the fact that $[\mathbf D_\mu,\mathbf D_\nu]=i \overline F_{\mu\nu}^3$, leads to the subsidiary condition
\begin{eqnarray}\label{subs}
\mathbf D^\nu \rho_\nu = 0,
\end{eqnarray}
which allows us to rewrite the equation as
\begin{equation}
\mathbf{D}_\mu^2 \rho_\nu  - 2 i \overline F_{\mu\nu}^3 \rho^\mu - m^2 \rho_\nu = 0.
\end{equation}
The condition \eqref{subs} and its conjugate are nothing else than the covariant (w.r.t.~the electromagnetic background) generalizations of the usual Proca subsidiary conditions $\partial^\nu \rho_\nu^\pm=0$.\\

\begin{figure}[t]
  \centering
  \scalebox{0.7}{
  \includegraphics{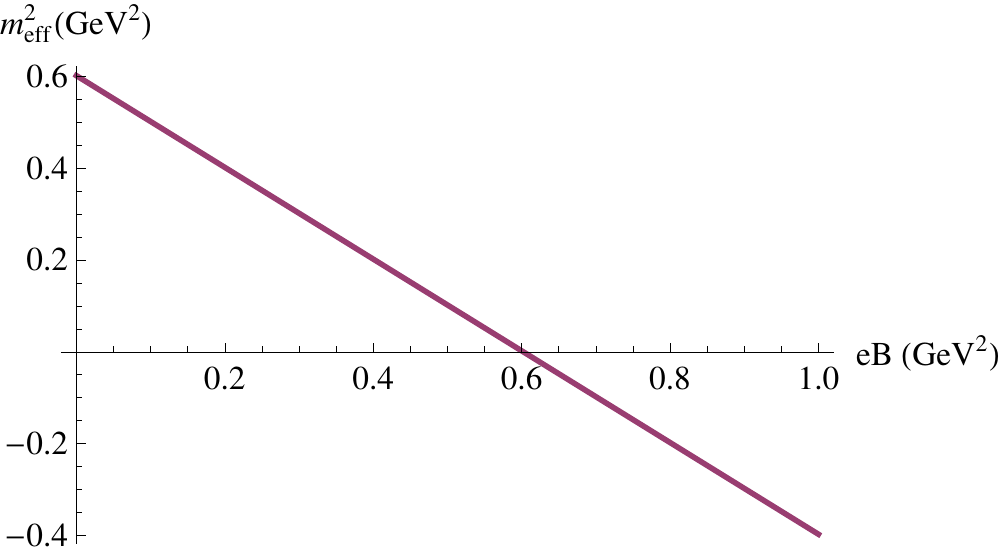}}
  \caption{The effective mass squared $\mathit{m_{eff}}^2 = m_\rho^2 - eB$ of the field combinations $\rho$ and $\rho^\dagger$ as a function of $eB$, for the case of antipodal embedding of the flavour branes. $\mathit{m_{eff}}^2$ goes through zero at $eB_c$.}
\label{meffantipodal}
\end{figure}

Fourier transforming $\rho_\nu \rightarrow e^{i (\vec k \cdot \vec x - E t )} \rho_\nu$, we find
that the transverse combinations\footnote{Because $\overline F_{\mu\nu}^{3}$ is only non-zero for $\mu,\nu \neq 0,3$ there is no magnetic moment coupling term for the longitudinal components of the $\rho$ fields, resulting in only the transverse components condensing according to the Landau equations of motion.} $\rho^-_1 \pm i \rho^-_2$ respectively get a negative or positive contribution $\mp eB$ to their effective mass squared as a consequence of their magnetic moment coupling to the magnetic field:
\begin{align}
E^2 (\rho^-_1 \pm i \rho^-_2)
&=  \left[ (k_2 + x_1 eB)^2 + k_3^2 - \partial_1^2 + m_\rho^2 \mp 2 eB \right] (\rho^-_1 \pm i \rho^-_2)  \nonumber\\
&=  \left[ - \partial_1^2 + (eB)^2 \left( x_1 + \frac{k_2}{eB}\right)^2 + k_3^2 + m_\rho^2 \mp 2 eB \right] (\rho^-_1 \pm i \rho^-_2)  \nonumber\\
&=  \left[ eB(2N+1) + k_3^2 + m_\rho^2 \mp 2 eB \right] (\rho^-_1 \pm i \rho^-_2).
\end{align}
The following combinations of the transverse (w.r.t.~$\vec{B}$) polarizations of the charged spin-1 $\rho$ meson fields,
\begin{eqnarray}\label{fieldcom}
\rho = \rho^-_1 + i \rho^-_2 \quad \mbox{ and }\quad  \rho^\dagger =  \rho_1^+ - i \rho_2^+,
\end{eqnarray}
have their spin aligned with the background magnetic field, $s_3 = 1$, which decreases their energy with an amount $eB$. As a consequence, the energy or effective mass of the fields $\rho$ and $\rho^\dagger$ in the lowest energy state, i.e.~in the lowest Landau level $N=0$ and with zero momentum $k_3$ along the direction of the magnetic field,
\begin{eqnarray} E =  \mathit{m_{eff}} = \sqrt{m_\rho^2 - eB},  \end{eqnarray}
becomes imaginary when the magnetic field reaches the critical value of $m_\rho^2$,
\begin{eqnarray}\label{crit} eB_c = m_\rho^2 \approx 0.60 \mbox{ GeV}^2,  \end{eqnarray}
see Figure \ref{meffantipodal}.  
We conclude from this that the field combinations  \eqref{fieldcom}
should experience a condensation, in accordance with \cite{Chernodub:2010qx,Chernodub:2011mc}.\\

Before concluding this Section, we mention that in principle there are also equations of motion for $\tilde A_u^a$, which, upon still ignoring the $F^4$ terms in the action, read (in the gauge $A_u=0$)
\begin{eqnarray}
\partial_u(\partial_\mu \omega^\mu) = \partial_u(\partial_\mu \rho^{\mu 0}) = \partial_u(\mathbf D_\mu \rho^{\mu -}) = \partial_u(\mathbf D_\mu^\dagger \rho^{\mu +}) = 0.
 \end{eqnarray}
The latter equations are automatically fulfilled due to the Proca subsidiary conditions for the neutral mesons, the subsidiary condition \eqref{subs} and its complex conjugate.

\subsection[Non-antipodal embedding]{Non-antipodal embedding ($u_0>u_K$): Chiral magnetic catalysis $\Rightarrow  eB_c > m_\rho^2$} \label{Landaunonantipodal}

Considering the case of non-antipodal embedding, $u_0 > u_K$, corresponds to taking into account effects of the constituents of the mesons in the analysis, as the distance between $u_0$ and $u_K$ is related to the constituent quark mass through (\ref{constmass}).
In contrast to the antipodal case, here the embedding of the flavour branes \'is affected by the magnetic field, as depicted in Figure \ref{changedembedding}.
The $eB$-dependent change in embedding models two effects which both will have an influence on the mass of the $\rho$ meson.
Firstly, the rise of $u_0$ and consequently of the constituent quark masses $m_u$ and $m_d$ with the magnetic field  models magnetic catalysis of chiral symmetry breaking. As the constituents get heavier it is expected that the $\rho$  meson will too, increasing the value of the critical magnetic field for condensation.
Secondly, the fact that the up- and down brane are separated, models the explicit breaking of the global chiral symmetry $U(2) \stackrel{eB}{\rightarrow} U(1)_u \times U(1)_d$. The split between the branes generates another mass mechanism for the charged $\rho$ mesons through a magnetically induced holographic Higgs mechanism: the charged $u \overline d$ and $d \overline u$ combinations correspond to strings with one endpoint on the up-brane and the other endpoint on the down-brane, these will now stretch between the separated branes and because of their string tension will develop a (stringy) mass term.

\begin{figure}[t]
  \centering
  \scalebox{0.7}{
  \includegraphics{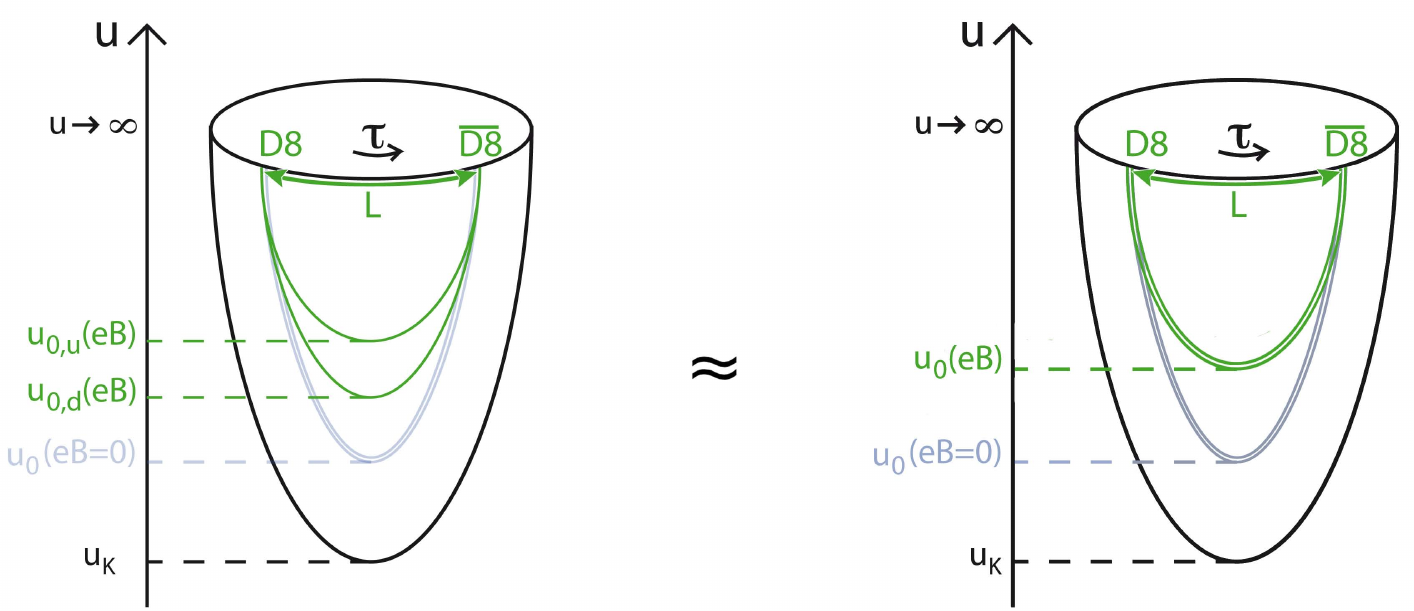}}
  \caption{The approximated $eB$-dependent embedding of the flavour branes, taking into account chiral magnetic catalysis but postulating coincident branes.}
\label{approxEmbedding}
\end{figure}

\begin{figure}[h!]
  \centering
  \scalebox{0.7}{
  \includegraphics{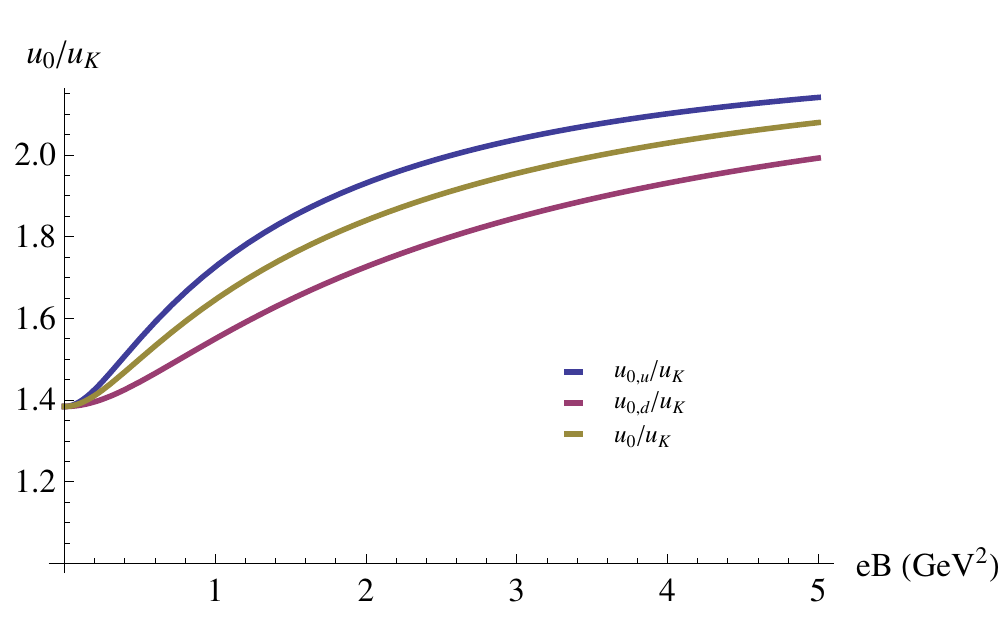}}
  \caption{The average $u_0(eB)$ (yellow) in the coincident branes approximation compared to $u_{0,u}(eB)$ (blue) and $u_{0,d}(eB)$ (red).}
\label{approxEmbeddingu0}
\end{figure}

Taking into account the splitting of the branes, $\overline \tau \not \sim \mathbf 1$, severely complicates the analysis, mainly because the evaluation of the STr in the action (\ref{detscalarplusvectorT}) becomes quite technical in the case of non-coincident branes (the STr no longer reduces to a normal Tr). We are in the current paper moreover particularly interested in the effect of the chiral magnetic catalysis on the $\rho$ meson mass and the critical magnetic field.
For this reason we will here approximate the embedding in Figure \ref{changedembedding} by an embedding where the branes are still coincident in the presence of the magnetic field ($\p_u \overline \tau \neq 0$ but $\overline \tau \sim \mathbf 1$), joining at an averaged value $u_0$ of the holographic radius, determined from
\begin{equation} \label{u0Baverage}
L_{average}(u_0,eB) = \left\{ (\ref{LconfifvB}) \text{ with every $A$ replaced by $A_{average} = \frac{(\sqrt{A_u} + \sqrt{A_d})^2}{2}$} \right\}  = L(eB=0)  \Rightarrow u_0(eB),
\end{equation}
see Figure \ref{approxEmbedding} and \ref{approxEmbeddingu0}.

This form of $L_{average}(u_0,eB)$ is the one you obtain when you postulate that the branes remain coincident in the presence of the magnetic field by using $g^{D8} = g^{D8} \mathbf 1$ instead of \eqref{metricnoncoincident}.
The same approximation, i.e.~not taking into account the splitting, is implicitly done when applying the magnetic field holographically by assigning a non-zero value only to $\overline A_\mu^3$ and not $\overline A_\mu^0$ (instead of (\ref{Aachtergrond})), as is quite often done in the literature, e.g.~in \cite{Bergman:2008sg,Ammon:2011je,Bu:2012mq}. In this approximation the Higgs effect on the $\rho$ meson mass is ignored, but we will elaborate on this in a follow-up paper where we will handle the non-approximated case of splitted branes rigorously by suitably adapting tools developed in \cite{Hashimoto:1997gm,Denef:2000rj}.

In the case $u_0>u_K$ and the current approximation of coincident branes we have
\begin{align}
\partial_u \overline \tau &= \sqrt{ \left(\frac{R}{u}\right)^{3} \frac{1}{f^2} \frac{u_0^8 f_0 A_0}{u^8 f A - u_0^8 f_0 A_0}}  \times \theta(u - u_0) \\
 \overline \tau &\sim \mathbf{1} \Rightarrow [A_r,\overline \tau] = 0  \text{ and $G_{uu}(\partial_u \overline \tau)$ drops out of the STr argument},
 \end{align}
simplifying (\ref{detscalarplusvectorT}) to a part in the scalar fluctuations and a part quadratic in the gauge fluctuations, given by
\begin{align}
\text{STr} & \hspace{1mm} e^{-\phi} \sqrt{-\det a} |_{\tilde A^2,T^{-2}}
= \text{Tr}  \hspace{1mm} e^{-\phi} g_{11}^2 \sqrt{G_{uu}} g_{S_4}^2 T^{-2}  \times  \nonumber\\ &
\left\{
\overline F_{12} g_{11}^{-2} \tilde F_{21}
- \overline F_{12} g_{11}^{-2} [\tilde A_1,\tilde A_2]
- \frac{1}{4} g_{\mu\mu}^{-1} g_{\nu\nu}^{-1} \tilde F_{\mu\nu}^2 - \frac{1}{2} g_{\mu\mu}^{-1} G_{uu}^{-1} \tilde F_{\mu u}^2 \right\},
\end{align}
where we again chose the gauge $A_u=0$ and only retained the lowest mesons of the meson towers in the fluctuation expansions (\ref{expansionvectorpion}) for the gauge field and $\tilde \tau(x^\mu,u) = \sum_n U^{(n)}(x^\mu) \phi_n(u)$ for the scalar field:
\begin{equation}
\tilde A_\mu^a = \rho_\mu^a(x^\mu) \psi(u) \text{ and } \tilde \tau = a_0(x^\mu) \phi(u) \quad (a=1,2) .
\end{equation}
Because both $\psi(z) \equiv \psi(u(z))$ and $\phi(z) \equiv \phi(u(z))$ (with $u(z) = u_0^3 + u_0 z^2$)
are even functions, the last line of (\ref{detscalarplusvectorT}) consists of integrals of the form $\int_{-\infty}^{\infty} dz  \{ \text{odd function of $z$} \} = 0$ and hence disappears.\\

\begin{figure}[h!]
  \hfill
  \begin{minipage}[t]{.45\textwidth}
    \begin{center}
      \scalebox{0.7}{
  \includegraphics{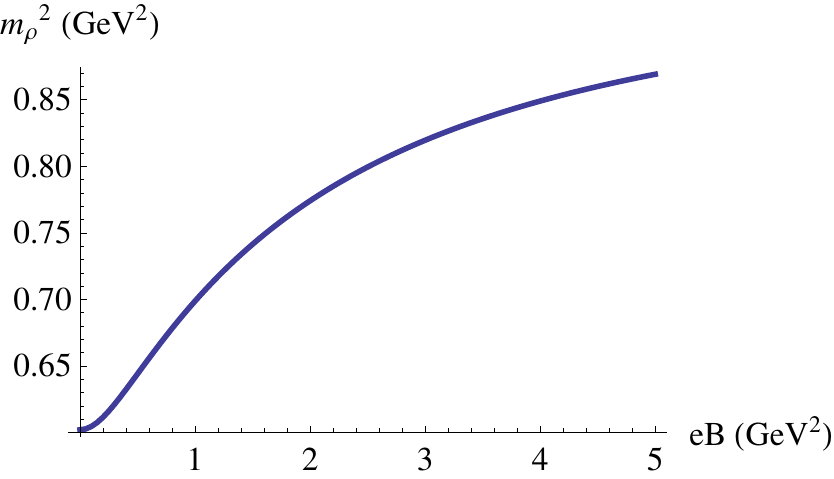}}
    \end{center}
  \end{minipage}
  \hfill
  \begin{minipage}[t]{.45\textwidth}
    \begin{center}
      \scalebox{0.7}{
  \includegraphics{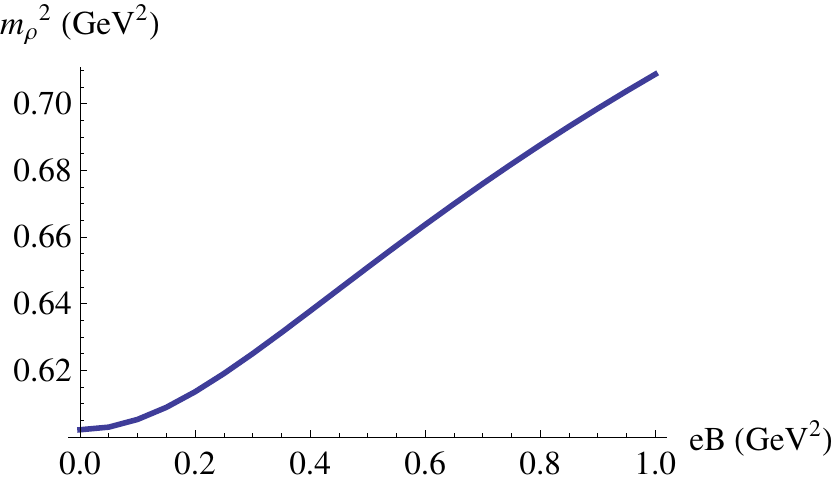}}
    \end{center}
  \end{minipage}
      \caption{The $\rho$ meson mass eigenvalue as a function of the magnetic field.}
	\label{mrhoB}
  \hfill
\end{figure}

\begin{figure}[t]
  \centering
  \scalebox{0.7}{
  \includegraphics{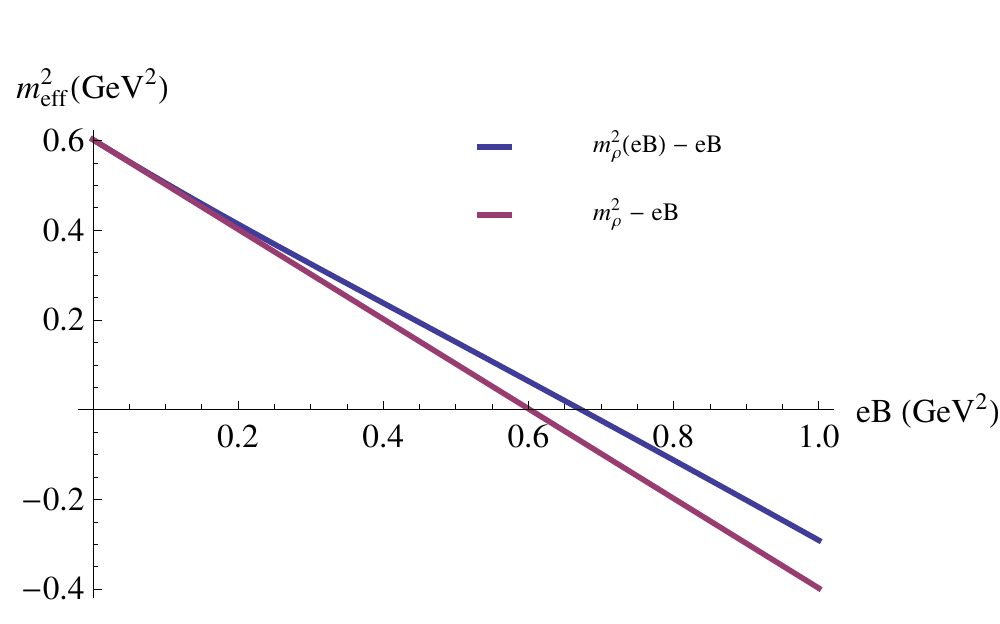}}
  \caption{The effective mass squared $\mathit{m_{eff}}^2 = m_\rho^2(eB) - eB$ of the field combinations $\rho$ and $\rho^\dagger$ as a function of $eB$, for the case of non-antipodal embedding of the flavour branes (blue), i.e. taking into account the effect of chiral magnetic catalysis through the $eB$-dependenve of $m_\rho$, compared to the case of antipodal embedding (red). $\mathit{m_{eff}}^2$ goes through zero at $eB_c$.}
\label{meffnonantipodal}
\end{figure}

The DBI-action to second order in the charged gauge field fluctuations can then again be written as
\begin{align}
S_{DBI} &= \int d^4 x \int du \sum_{a=1}^2  \left\{ -\frac{1}{4} f_1 (\tilde F_{\mu\nu}^a)^2 - \frac{1}{2} f_2 (\tilde F_{\mu u}^a)^2 - \frac{1}{2} f_3 \sum_{\mu,\nu=1}^2 \overline F_{\mu\nu}^3 \epsilon_{3ab} \tilde A^{\mu a} \tilde A^{\nu b} \right\} \nonumber\\
&=  \int d^4x \int du \sum_{a=1}^2 \left\{ -\frac{1}{4} f_1(\mathcal F_{\mu\nu}^a)^2 \psi^2 - \frac{1}{2} f_2(\rho_\mu^a)^2 (\partial_u \psi)^2 - \frac{1}{2} f_3 \sum_{\mu,\nu=1}^2 \overline F_{\mu\nu}^3 \epsilon_{3ab} \rho^{\mu a} \rho^{\nu b} \psi^2 \right\} + \text{pion action}
\end{align}
with the functions $f_i$ ($i=1..3$) dependent on both $u$ \'and $eB$ this time (through the $eB$-dependence of the embedding function $\partial_u \overline \tau$ in $G_{uu}$):
\begin{eqnarray}
f_1(u,eB) = f_3(u,eB)  &=& T_8 V_4 (2\pi\alpha')^2 e^{-\phi} g^2_{S_4} \sqrt{G_{uu}}\,, \\
f_2(u,eB) &=& T_8 V_4 (2\pi\alpha')^2 e^{-\phi} g^2_{S_4} g_{11} G_{uu}^{-1/2}.
\end{eqnarray}
In order to obtain a canonical kinetic term and mass term for the $\rho$ mesons in the effective four-dimensional action, we demand the $\psi(u)$ to fulfill the standard conditions
\begin{align}
\int_{u_0(eB)}^\infty du \hspace{1mm} f_1(u,eB) \psi^2 &= 1\,, \label{Bnormalization}\\
\int_{u_0(eB)}^\infty du \hspace{1mm} f_2(u,eB) (\partial_u \psi)^2 &= m_\rho^2(eB).
\end{align}
These conditions combine to the eigenvalue equation (\ref{finiteTeigwvgl}) with $\gamma =\gamma_{\braket{eB}} = (\ref{gammaB})$ with every $A$ replaced by $A_{average} = \frac{(\sqrt{A_u} + \sqrt{A_d})^2}{2}$:
\begin{equation} \label{Beigvproblem}
-u^{1/2} \gamma_{\braket{eB}}^{-1/2} \partial_u (u^{5/2} \gamma_{\braket{eB}}^{-1/2} \partial_u \psi) = R^3 m_\rho^2(eB) \psi.
\end{equation}
Numerically solving this eigenvalue equation for $m_\rho^2(eB)$ (see the Appendix for details) boils down to taking the averaged magnetic catalysis of chiral symmetry breaking into account, as we include the effect of $eB$ on the embedding of the flavour branes, both through the induced metric on the branes and the changed value of $u_0$. The resulting $\rho$ meson mass eigenvalue $m_\rho^2(eB)$ as a function of $eB$ is depicted in Figure \ref{mrhoB}. As expected on grounds of Figure \ref{mqfig}, it is an increasing function of magnetic field (where it should be noted that working in the approximation of coincident branes means that we consider the effect of an averaged increase for $m_q(eB)$, equal for up and down, rather than the exact $eB$-dependences in Figure \ref{mqfig}).
The corresponding eigenfunction $\psi$ fulfilling the boundary conditions $\psi'(z=0) = 0$ and $\psi(z\rightarrow\pm\infty)=0$ can be used to evaluate the last integral over $u$ in the above action, which determines the gyromagnetic coupling constant $k$ to be one,
\begin{equation}
\int_{u_0(eB)}^\infty du \hspace{1mm} f_3(u,eB) \psi^2 = k = 1,
\end{equation}
or $g=k+1=2$, again because of $f_1 = f_3$. The effective four-dimensional action thus again takes the form of the standard four-dimensional action used to describe the coupling of charged vector mesons to an external magnetic field (i.e. the Proca action \cite{Obukhov:1984xb,Tsai:1972iq} or DSGS action for self-consistent $\rho$ meson quantum electrodynamics to second order in the fields \cite{Djukanovic:2005ag}), but with the influence of the constituent quarks reflected in the magnetic field dependence of $m_\rho$: 
\begin{equation}
S_{eff} = \int d^4x \sum_{a=1}^2 \left\{ -\frac{1}{4}(\mathcal F_{\mu\nu}^a)^2 - \frac{1}{2} m_\rho^2(eB) (\rho_\mu^a)^2  - \frac{1}{2} \sum_{\mu,\nu=1}^2 \overline F_{\mu\nu}^3 \epsilon_{3ab} \rho^{\mu a} \rho^{\nu b} \right\}.
\end{equation}
Completely analogous to the derivation in the preceding subsection, this action describes that the $\rho$ and $\rho^\dagger$ fields in the lowest Landau energy level will get an effective mass
\begin{eqnarray} E = \mathit{m_{eff}} = \sqrt{m_\rho^2(eB) - eB},  \end{eqnarray}
which becomes imaginary when $eB_c = m_\rho^2(eB_c)$, at 
\begin{eqnarray}\label{crit} eB_c = m_\rho^2(eB_c) \approx 0.67 \mbox{ GeV}^2 \approx 1.1 \, m_ \rho^2,   \end{eqnarray}
see Figure \ref{meffnonantipodal}. 
The averaged chiral magnetic catalysis pushes the critical magnetic field for condensation to a higher value, as expected. The increase is of the order of 10 percent. In future work, we shall probe the further increase in $eB_c$ due to the already alluded to Higgs effect caused by the flavour brane separation, to investigate to what extent we can get closer to the lattice estimate for the critical magnetic field, i.e. $eB_c$ of the order of 1 GeV$^2$ \cite{Braguta:2011hq}.

\subsection{The Chern-Simons action and mixing with pions}\label{CSpion}
In principle, the DBI action \eqref{nonabelian} needs to be complemented with a Chern-Simons piece which serves as the chiral anomaly in the QCD-like boundary theory:
\begin{eqnarray}
S_{CS} = \frac{N_c}{24 \pi^2} i \int \text{Tr} \left(A F^2 - \frac{1}{2} A^3 F + \frac{1}{10} A^5 \right),
\end{eqnarray}
where the notation implies wedge products of differential forms. Since it is a factor $\lambda$ smaller than the DBI-action and $\lambda \approx 15 \gg 1$, we have ignored the CS-action in our above analysis.\\

When one applies a magnetic field, one might intuitively guess that a quark and anti-quark forming a pion bound state might try to align their spins, thereby transforming into a $\rho$ meson. In field theoretic terms, one might therefore expect a mixing between pions and $\rho$ mesons due to a magnetic field. Given the different intrinsic parity of the participants in this mixing, this would necessarily correspond to an anomaly driven process, i.e.~related to the Chern-Simons piece of the complete Sakai-Sugimoto action. It therefore looks interesting to investigate the CS-action to order meson gauge field squared in some more detail. We find that, after renormalizing by subtracting appropriate boundary terms \cite{Bergman:2008qv}, the action
\begin{align*}
S_{CS} &\sim  \int \text{Tr} \left( \epsilon^{mnpqr} A_m F_{np} F_{qr} + \mathcal O(\tilde A^3) \right) \\
	     &\sim  \int 4 \text{Tr} \left( \tilde A_3 \overline{F}_{21} \tilde F_{04} + \tilde A_0 \overline{F}_{21} \tilde F_{43} - \tilde A_3 \tilde F_{40}\overline{F}_{21} - \tilde A_0 \tilde F_{43} \overline{F}_{12} \right.\\
 & \qquad \qquad \qquad \left. + \overline{A}_2 [\tilde F_{43}\tilde F_{10} + \tilde F_{10}\tilde F_{43} + \tilde F_{41}\tilde F_{03} +\tilde F_{03}\tilde F_{41} +\tilde F_{31}\tilde F_{40} +\tilde F_{40}\tilde F_{31}] \right),
\end{align*}
does contribute interesting terms of the form $\rho \pi B$ (see also \cite{Sakai:2005yt}),
\begin{eqnarray}  \sim B \int \left\{
\partial_{[0} \pi^0 \rho_{3]}^0 + \frac{1}{2} \left(\partial_{[0} \pi^+ \rho_{3]}^-  + \partial_{[0} \pi^- \rho_{3]}^+\right) \right\},
\end{eqnarray}
indeed describing couplings between $\rho$-mesons and pions (and other axial mesons that we have not taken into account). However, since there are no direct couplings between the transverse components ($\mu = 1,2$) of the $\rho$ fields and pions, we can conclude that, at the level of a $\mathcal O(\tilde A^2)$ analysis, taking the CS-action into account will not affect the presence of the $\rho$ meson condensate which is related to the transverse field components, see eq.~\eqref{fieldcom}.

\section{Summary and outlook}\label{conc}

We studied the $\rho$ meson mass dependence on an external constant magnetic field, $\vec{B}=B\vec{e}_3$, using the $N_f=2$ Sakai-Sugimoto model, which is a holographic dual of a QCD-like theory in the quenched approximation and chiral limit. We fixed all free parameters by matching them to known QCD values at zero magnetic field. As a first main result, we found that two non-antipodal embedded flavour branes do not respond equally strong to the magnetic field, which is the geometric translation of the up- and down-quark coupling to the magnetic field with different electromagnetic charges (respectively $\frac{2}{3}e$ and $-\frac{1}{3}e$), resulting in the up- and down-brane becoming separated at $eB\neq0$. This means that the chiral magnetic catalysis is  stronger for the up-quarks than for the down-quarks. The presence of the magnetic field thus induces an explicit breaking of the chiral symmetry, which manifests itself in the non-Abelian DBI-action for the flavour branes reducing to the sum of two Abelian actions, one for the up- and one for the down-brane. As a result of the chiral magnetic catalysis, the constituent quarks gain mass (the up-quarks thus getting heavier than the down-quarks), and consequently also the $\rho$ meson mass eigenvalue gets influenced. \\

As a second main result, we have shown that the effective action for the $\rho$ mesons in the presence of a magnetic background leads to Landau levels, whereby the effective masses of the combinations $\rho_1^\pm \mp  i\rho_2^\pm$ of transverse components of the $\rho$ meson become tachyonic, indicative of a condensation of these charged fields. The value for the critical magnetic field at the onset of the condensation was estimated, taking into account the contribution to the effective $\rho$ meson mass
due to an averaged (i.e. equal for up and down) chiral magnetic catalysis, as $eB_c \approx 1.1\,m_\rho^2 \approx 0.67~\text{GeV}^2$. Such a $\rho$ meson condensation effect at very strong magnetic field was predicted and later confirmed on the lattice in the papers \cite{Chernodub:2010qx,Chernodub:2011mc,Braguta:2011hq}.
We will present the effect of the splitting of the branes on the $\rho$ meson mass (i.e.~the effect of the full non-averaged chiral magnetic catalysis plus a stringy Higgs effect) as well as the discussion of stability in the scalar meson sector in a follow-up paper \cite{nele3}. \\

Another step in this research program will be the explicit construction of the new vacuum in the condensed phase (either in the case of coincident or separated branes), to see to what extent the findings of  \cite{Chernodub:2010qx,Chernodub:2011mc,Braguta:2011hq} can be further corroborated by holographic tools. Therefore, we shall need to discuss the condensation of $\rho_1^\pm \mp i\rho_2^\pm$ along the lines of \cite{Chernodub:2011gs,Bu:2012mq}. It would also be interesting to include temperature effects, to study how the (super)conducting condensed $\rho$ meson phase evolves. Hopefully, once these tasks are carried out, we shall be able to tell something more about a possible $(T,eB)$ phase diagram of QCD as sketched in \cite{Chernodub:2011tv}. In principle, using the tools developed in this paper for the $\rho$ mesons, also the axial vector mesons can be investigated. It would be interesting to find out whether these could condense or not, as they also couple to the magnetic field, both directly and indirectly. They were conjectured not to condense in \cite{Chernodub:2011mc}. Another line of investigation concerns the chiral restoration and deconfinement temperature structure of the $N_f=2$ Sakai-Sugimoto model in the presence of the magnetic field. This will be presented elsewhere \cite{nele2}.

\appendix

\section{Solving the mass eigenvalue equation}

We numerically solve the eigenvalue equation (\ref{finiteTeigwvgl}) with normalization condition \eqref{orthonormconditiepsi}. Since
\begin{equation}\label{extra3}
u^{-1/2}\gamma^{1/2}\sim u^{-1/2} \qquad\text{for } u\sim \infty,
\end{equation}
normalizability requires that
\begin{equation}\label{extra4}
    \psi_n(u=\infty)=0.
\end{equation}
For numerical purposes, it is easier to work on a compact interval. The following transformation
\begin{equation}\label{extra5}
    \frac{u^3}{u_0^3}=\frac{1}{\cos^2x}
\end{equation}
maps  the interval under investigation onto $x\in[-\pi/2;\pi/2]$, with the boundary conditions \eqref{extra4} now reading
\begin{equation}\label{mass11}
    \psi_n(\pm \pi/2)=0.
\end{equation}
The differential equation \eqref{finiteTeigwvgl} transforms into
\begin{equation}\label{extra6}
- \frac{9}{4} u_0 \frac{\cos^4x}{\sin x}    \sqrt{\frac{1-\frac{u_K^3}{u_0^3}\cos^2x}{\cos^{16/3}x}-f_0}~\frac{\p}{\p x}\left(\frac{\cos^{8/3}x}{\sin x}    \sqrt{\frac{1-\frac{u_K^3}{u_0^3}\cos^2x}{\cos^{16/3}x}-f_0}~\frac{\p}{\p x}\right)\psi_n(x)=R^3\lambda_n\psi_n(x),
\end{equation}
where we denoted the mass eigenvalue $m_n^2$ with $\lambda_n$. Due to the reflection symmetry $x\to-x$ we can split up the eigenfunction set in even/odd $\psi_n(x)$'s and focus on the interval $[0,\pi/2]$. Analogously as explained in \cite{Sakai:2004cn,Peeters:2006iu}, the even/odd eigenfunctions correspond to odd/even parity mesons. We can thus demand that
\begin{equation}\label{mass10}
    \psi_n(0)=0\qquad\text{or}\qquad\p_x\psi_n(0)=0.
\end{equation}
For the $\rho$ meson, we must look at the odd parity sector, in particular the even eigenfunction with the lowest eigenvalue.  We can temporarily replace \eqref{finiteTeigwvgl} and associated boundary conditions with the initial value problem
\begin{equation}\label{mass16}
- u_0 \frac{\cos^4x}{\sin x}    \sqrt{\frac{1-\frac{u_K^3}{u_0^3}\cos^2x}{\cos^{16/3}x}-f_0}~\frac{\p}{\p x}\left(\frac{\cos^{8/3}x}{\sin x}    \sqrt{\frac{1-\frac{u_K^3}{u_0^3}\cos^2x}{\cos^{16/3}x}-f_0}~\frac{\p}{\p x}\right)\psi=\frac{1}{M^3}\Lambda\psi\,,\psi(0)=1\,,\psi'(0)=0
\end{equation}
where $\Lambda$ is treated as a ``shooting'' parameter. We numerically solved the previous differential equation for each value of $\Lambda$ to give a unique $\psi_\Lambda(x)$, consistent with the initial conditions. Since the coefficient functions appearing in \eqref{mass16} display a delicate behaviour $\sim 0/0$ around $x=0$, some care is needed when using a numerical package, as this is precisely where we are imposing our initial conditions. The solution around $x=0$ can however be easily obtained using a Taylor expansion, and fed into the numerical procedure. Once the $\psi_\Lambda(x)$ is known, we can solve the equation
\begin{equation}\label{mass17}
    F^{\text{even}}(\Lambda)\equiv\psi_\Lambda(\pi/2)=0
\end{equation}
for $\Lambda$, which is then precisely the mass eigenvalue of the original eigenvalue problem.\\

For the $eB$-dependent eigenvalue problem (\ref{Beigvproblem}) with normalization condition (\ref{Bnormalization}), the corresponding initial value problem can be written as (\ref{mass16}) with every $f_0 \rightarrow f_0 A_0/A$,  where in the case of the approximation of coincident branes $A$ is replaced by $A_{average} = \frac{(\sqrt{A_u} + \sqrt{A_d})^2}{2}$, and where it is understood that each $u_0$ appearing in the eigenvalue equation is now $eB$-dependent, see $u_0(eB)$ from (\ref{u0Baverage}). This can accordingly be solved numerically, using the explained shooting method, for the $eB$-dependent mass eigenvalue $\Lambda(eB) = m_\rho^2(eB)$.

\section*{Acknowledgements}
We wish to thank M.~N.~Chernodub, A.~J.~Mizher and J.~Van Doorsselaere for interesting discussions, T.~Cohen for bringing the explicit chiral symmetry breaking to our attention,  and S.~Callebaut for a helping hand in drawing the figures. N.~Callebaut and D.~Dudal are supported by the Research Foundation Flanders (FWO Vlaanderen).


\begin{thebibliography}{99}
\bibitem{:2009uh}
B.~I.~Abelev {\it et al.} [ STAR Collaboration ],
  Phys.\ Rev.\ Lett.\  {\bf 103 } (2009)  251601.
  [arXiv:0909.1739 [nucl-ex]].

\bibitem{Skokov:2009qp}
  V.~Skokov, A.~Y.~Illarionov, V.~Toneev,
  Int.\ J.\ Mod.\ Phys.\  {\bf A24 } (2009)  5925-5932.
  [arXiv:0907.1396 [nucl-th]].

\bibitem{Deng:2012pc}
W.~-T.~Deng and X.~-G.~Huang,
  Phys.\ Rev.\ C {\bf 85} (2012) 044907
  [arXiv:1201.5108 [nucl-th]].

\bibitem{Bzdak:2011yy}
A.~Bzdak and V.~Skokov,
  Phys.\ Lett.\ B {\bf 710} (2012) 171
  [arXiv:1111.1949 [hep-ph]].

\bibitem{Tuchin:2013ie}
  K.~Tuchin,
 arXiv:1301.0099 [hep-ph].  


\bibitem{Schramm:1991ex}
  S.~Schramm, B.~Muller and A.~J.~Schramm,
  Mod.\ Phys.\ Lett.\ A {\bf 7} (1992) 973.


\bibitem{Fukushima:2008xe}
  K.~Fukushima, D.~E.~Kharzeev, H.~J.~Warringa,
  Phys.\ Rev.\  {\bf D78 } (2008)  074033.
  [arXiv:0808.3382 [hep-ph]].



\bibitem{Kharzeev:2010gd}
  D.~E.~Kharzeev, H.~-U.~Yee,
  Phys.\ Rev.\  {\bf D83 } (2011)  085007.
  [arXiv:1012.6026 [hep-th]].

\bibitem{Basar:2010zd}
  G.~Basar, G.~V.~Dunne, D.~E.~Kharzeev,
  Phys.\ Rev.\ Lett.\  {\bf 104 } (2010)  232301.
  [arXiv:1003.3464 [hep-ph]].

\bibitem{Fukushima:2010fe}
  K.~Fukushima, M.~Ruggieri, R.~Gatto,
  Phys.\ Rev.\  {\bf D81 } (2010)  114031.
  [arXiv:1003.0047 [hep-ph]].

\bibitem{Fraga:2008qn}
  E.~S.~Fraga, A.~J.~Mizher,
  Phys.\ Rev.\  D {\bf 78} (2008) 025016.
  [arXiv:0804.1452 [hep-ph]].

\bibitem{Mizher:2010zb}
  A.~J.~Mizher, M.~N.~Chernodub, E.~S.~Fraga,
  Phys.\ Rev.\  D {\bf 82} (2010) 105016.
  [arXiv:1004.2712 [hep-ph]].

\bibitem{Mizher:2011wd}
A.~J.~Mizher, E.~S.~Fraga, M.~N.~Chernodub,  PoS(FacesQCD) (2010)  020 [arXiv:1103.0954 [hep-ph]].

\bibitem{Gatto:2010pt}
  R.~Gatto, M.~Ruggieri,
  Phys.\ Rev.\  {\bf D83 } (2011)  034016.
  [arXiv:1012.1291 [hep-ph]].

\bibitem{Gatto:2010qs}
  R.~Gatto, M.~Ruggieri,
  Phys.\ Rev.\  {\bf D82 } (2010)  054027.
  [arXiv:1007.0790 [hep-ph]].

\bibitem{Chernodub:2011fr}
  M.~N.~Chernodub and A.~S.~Nedelin,
  Phys.\ Rev.\ D {\bf 83} (2011) 105008
  [arXiv:1102.0188 [hep-ph]].

\bibitem{Frasca:2011zn}
  M.~Frasca and M.~Ruggieri,
  Phys.\ Rev.\ D {\bf 83} (2011) 094024
  [arXiv:1103.1194 [hep-ph]].

\bibitem{Kashiwa:2011js}
  K.~Kashiwa,
  Phys.\ Rev.\ D {\bf 83} (2011) 117901
  [arXiv:1104.5167 [hep-ph]].

\bibitem{Nam:2011vn}
  S.~-i.~Nam and C.~-W.~Kao,
  Phys.\ Rev.\ D {\bf 83} (2011) 096009
  [arXiv:1103.6057 [hep-ph]].

\bibitem{Andersen:2011ip}
  J.~O.~Andersen and R.~Khan,
  Phys.\ Rev.\ D {\bf 85} (2012) 065026
  [arXiv:1105.1290 [hep-ph]].



\bibitem{Andersen:2012bq}
J.~O.~Andersen and A.~Tranberg,
  JHEP {\bf 1208} (2012) 002
  [arXiv:1204.3360 [hep-ph]].

\bibitem{Andersen:2012jf}
J.~O.~Andersen and A.~A.~Cruz,
  arXiv:1211.7293 [hep-ph].



\bibitem{Fukushima:2012xw}
K.~Fukushima and J.~M.~Pawlowski,
  Phys.\ Rev.\ D {\bf 86} (2012) 076013
  [arXiv:1203.4330 [hep-ph]].

\bibitem{Fukushima:2012kc}
K.~Fukushima and Y.~Hidaka,
  arXiv:1209.1319 [hep-ph].

\bibitem{Fraga:2012fs}
E.~S.~Fraga and L.~F.~Palhares,
  Phys.\ Rev.\ D {\bf 86} (2012) 016008
  [arXiv:1201.5881 [hep-ph]].

\bibitem{Fraga:2012ev}
E.~S.~Fraga, J.~Noronha and L.~F.~Palhares,
  arXiv:1207.7094 [hep-ph].

\bibitem{Blaizot:2012sd}
J.~-P.~Blaizot, E.~S.~Fraga and L.~F.~Palhares,
  arXiv:1211.6412 [hep-ph].



\bibitem{Simonov:2012if}
Y.~.A.~Simonov, B.~O.~Kerbikov and M.~A.~Andreichikov,
  arXiv:1210.0227 [hep-ph].

\bibitem{Simonov:2012mf}
Y.~.A.~Simonov,
  arXiv:1212.3118 [hep-ph].


\bibitem{Chernodub:2010qx}
  M.~N.~Chernodub,
  Phys.\ Rev.\  D {\bf 82} (2010) 085011
  [arXiv:1008.1055 [hep-ph]].

\bibitem{Chernodub:2011mc}
  M.~N.~Chernodub,
  Phys.\ Rev.\ Lett.\  {\bf 106 } (2011)  142003.
  [arXiv:1101.0117 [hep-ph]].

\bibitem{Chernodub:2011gs}
  M.~N.~Chernodub, J.~Van Doorsselaere and H.~Verschelde,
  Phys.\ Rev.\ D {\bf 85} (2012) 045002
  [arXiv:1111.4401 [hep-ph]].


\bibitem{Hidaka:2012mz}
Y.~Hidaka and A.~Yamamoto,
  arXiv:1209.0007 [hep-ph].

\bibitem{Chernodub:2012zx}
M.~N.~Chernodub,
  Phys.\ Rev.\ D {\bf 86} (2012) 107703
  [arXiv:1209.3587 [hep-ph]].


\bibitem{Chernodub:2012fi}
M.~N.~Chernodub, J.~Van Doorsselaere and H.~Verschelde,
  arXiv:1203.5963 [hep-ph].


\bibitem{Chernodub:2012mu}
M.~N.~Chernodub, J.~Van Doorsselaere, T.~Kalaydzhyan and H.~Verschelde,
  arXiv:1212.3168 [hep-ph].

\bibitem{Buividovich:2008wf}
  P.~V.~Buividovich, M.~N.~Chernodub, E.~V.~Luschevskaya, M.~I.~Polikarpov,
  Phys.\ Lett.\  {\bf B682 } (2010)  484-489.
  [arXiv:0812.1740 [hep-lat]].

\bibitem{Buividovich:2009wi}
  P.~V.~Buividovich, M.~N.~Chernodub, E.~V.~Luschevskaya, M.~I.~Polikarpov,
  Phys.\ Rev.\  {\bf D80 } (2009)  054503.
  [arXiv:0907.0494 [hep-lat]].

\bibitem{Buividovich:2010tn}
  P.~V.~Buividovich, M.~N.~Chernodub, D.~E.~Kharzeev, T.~Kalaydzhyan, E.~V.~Luschevskaya, M.~I.~Polikarpov,
  Phys.\ Rev.\ Lett.\  {\bf 105 } (2010)  132001.
  [arXiv:1003.2180 [hep-lat]].

\bibitem{Braguta:2011hq}
  V.~V.~Braguta, P.~V.~Buividovich, M.~N.~Chernodub and M.~I.~Polikarpov,
  Phys.\ Lett.\ B {\bf 718} (2012) 667
  [arXiv:1104.3767 [hep-lat]].


\bibitem{D'Elia:2010nq}
  M.~D'Elia, S.~Mukherjee, F.~Sanfilippo,
  Phys.\ Rev.\  D {\bf 82} (2010) 051501.
  [arXiv:1005.5365 [hep-lat]].

\bibitem{D'Elia:2011zu}
  M.~D'Elia and F.~Negro,
  Phys.\ Rev.\ D {\bf 83} (2011) 114028
  [arXiv:1103.2080 [hep-lat]].


\bibitem{Ilgenfritz:2012fw}
E.~-M.~Ilgenfritz, M.~Kalinowski, M.~Muller-Preussker, B.~Petersson and A.~Schreiber,
  Phys.\ Rev.\ D {\bf 85} (2012) 114504
  [arXiv:1203.3360 [hep-lat]].

\bibitem{Bali:2012zg}
G.~S.~Bali, F.~Bruckmann, G.~Endrodi, Z.~Fodor, S.~D.~Katz and A.~Schafer,
  Phys.\ Rev.\ D {\bf 86} (2012) 071502
  [arXiv:1206.4205 [hep-lat]].

\bibitem{Bali:2011qj}
G.~S.~Bali, F.~Bruckmann, G.~Endrodi, Z.~Fodor, S.~D.~Katz, S.~Krieg, A.~Schafer and K.~K.~Szabo,
  JHEP {\bf 1202} (2012) 044
  [arXiv:1111.4956 [hep-lat]].

\bibitem{Yamamoto:2011gk}
  A.~Yamamoto,
  Phys.\ Rev.\ Lett.\  {\bf 107} (2011) 031601
  [arXiv:1105.0385 [hep-lat]].

\bibitem{Johnson:2008vna}
C.~V.~Johnson, A.~Kundu, JHEP {\bf 0812} (2008) 053.  [arXiv:0803.0038 [hep-th]].

\bibitem{Zayakin:2008cy}
  A.~V.~Zayakin,
  JHEP {\bf 0807} (2008) 116.
  [arXiv:0807.2917 [hep-th]].

\bibitem{Yee:2009vw}
  H.~-U.~Yee,
  JHEP {\bf 0911 } (2009)  085.
  [arXiv:0908.4189 [hep-th]].

\bibitem{Callebaut:2011uc}
N.~Callebaut, D.~Dudal, H.~Verschelde, PoS(FacesQCD) (2010)  021.  [arXiv:1102.3103 [hep-ph]].

\bibitem{Preis:2010cq}
  F.~Preis, A.~Rebhan, A.~Schmitt,
  JHEP {\bf 1103 } (2011)  033.
  [arXiv:1012.4785 [hep-th]].

\bibitem{Rebhan:2008ur}
  A.~Rebhan, A.~Schmitt, S.~A.~Stricker,
  JHEP {\bf 0905 } (2009)  084.
  [arXiv:0811.3533 [hep-th]].

\bibitem{Gynther:2010ed}
  A.~Gynther, K.~Landsteiner, F.~Pena-Benitez, A.~Rebhan,
  JHEP {\bf 1102 } (2011)  110.
  [arXiv:1005.2587 [hep-th]].

\bibitem{Bergman:2008sg}
  O.~Bergman, G.~Lifschytz, M.~Lippert,
  JHEP {\bf 0805 } (2008)  007.
  [arXiv:0802.3720 [hep-th]].

\bibitem{Ammon:2011je}
M.~Ammon, J.~Erdmenger, P.~Kerner and M.~Strydom, Phys.\ Lett.\ B {\bf 706} (2011) 94.  [arXiv:1106.4551 [hep-th]].

\bibitem{Bu:2012mq}
Y.~-Y.~Bu, J.~Erdmenger, J.~P.~Shock and M.~Strydom. [arXiv:1210.6669 [hep-th]]


\bibitem{Filev:2011mt}
V.~G.~Filev and D.~Zoakos,
  JHEP {\bf 1108} (2011) 022
  [arXiv:1106.1330 [hep-th]].


\bibitem{Erdmenger:2011bw}
J.~Erdmenger, V.~G.~Filev and D.~Zoakos,
  JHEP {\bf 1208} (2012) 004
  [arXiv:1112.4807 [hep-th]].


\bibitem{Evans:2010xs}
  N.~Evans, T.~Kalaydzhyan, K.~-y.~Kim, I.~Kirsch,
  JHEP {\bf 1101 } (2011)  050.
  [arXiv:1011.2519 [hep-th]].

\bibitem{boek}
D.~E.~Kharzeev, K.~Landsteiner, A.~Schmitt and H.~-U.~Yee, 
   and references therein, to appear in Lecture Notes in Physics on
``Strongly interacting matter in magnetic fields'', arXiv:1211.6245 [hep-ph].

\bibitem{Donos:2011pn}
  A.~Donos, J.~P.~Gauntlett and C.~Pantelidou,
  Class.\ Quant.\ Grav.\  {\bf 29} (2012) 194006
  [arXiv:1112.4195 [hep-th]].

\bibitem{Almuhairi:2011ws}
  A.~Almuhairi and J.~Polchinski,
  arXiv:1108.1213 [hep-th].

\bibitem{Vachaspati:1991nm}
  T.~Vachaspati,
  Phys.\ Lett.\  {\bf B265 } (1991)  258-261.

\bibitem{Ambjorn:1989sz}
  J.~Ambjorn, P.~Olesen,
  Int.\ J.\ Mod.\ Phys.\  {\bf A5 } (1990)  4525-4558.


\bibitem{VanDoorsselaere:2012zb}
J.~Van Doorsselaere,
  arXiv:1206.6205 [hep-ph].


\bibitem{Sakai:2004cn}
  T.~Sakai, S.~Sugimoto,
  Prog.\ Theor.\ Phys.\  {\bf 113} (2005) 843.
  [arXiv:hep-th/0412141].


\bibitem{Sakai:2005yt}
  T.~Sakai, S.~Sugimoto,
  Prog.\ Theor.\ Phys.\  {\bf 114} (2005) 1083.
  [arXiv:hep-th/0507073].

\bibitem{Kinar:1998vq}
  Y.~Kinar, E.~Schreiber and J.~Sonnenschein,
  Nucl.\ Phys.\ B {\bf 566} (2000) 103.
  [hep-th/9811192].


\bibitem{Maldacena:1997re}
  J.~M.~Maldacena,
  Adv.\ Theor.\ Math.\ Phys.\  {\bf 2 } (1998)  231-252.
  [hep-th/9711200].


\bibitem{Aharony:2006da}
  O.~Aharony, J.~Sonnenschein and S.~Yankielowicz,
  Annals Phys.\  {\bf 322} (2007) 1420
  [arXiv:hep-th/0604161].


\bibitem{Burrington:2007qd}
  B.~A.~Burrington, V.~S.~Kaplunovsky, J.~Sonnenschein,
  JHEP {\bf 0802 } (2008)  001.
  [arXiv:0708.1234 [hep-th]].

\bibitem{Miransky:2002rp}
  V.~A.~Miransky, I.~A.~Shovkovy,
  Phys.\ Rev.\  {\bf D66 } (2002)  045006.
  [hep-ph/0205348].

\bibitem{Shushpanov:1997sf}
  I.~A.~Shushpanov, A.~V.~Smilga,
  Phys.\ Lett.\  B {\bf 402} (1997) 351.
  [arXiv:hep-ph/9703201].

\bibitem{Bergman:2007pm}
  O.~Bergman, S.~Seki, J.~Sonnenschein,
  JHEP {\bf 0712 } (2007)  037.
  [arXiv:0708.2839 [hep-th]].


\bibitem{Tseytlin:1997csa}
  A.~A.~Tseytlin,
  Nucl.\ Phys.\  {\bf B501 } (1997)  41-52.
  [hep-th/9701125].

\bibitem{Myers:2003bw}
  R.~C.~Myers,
  Class.\ Quant.\ Grav.\  {\bf 20} (2003) S347
  [hep-th/0303072].


\bibitem{Howe:2006rv}
  P.~S.~Howe, U.~Lindstrom and L.~Wulff,
  JHEP {\bf 0702} (2007) 070
  [hep-th/0607156].


\bibitem{Wulff:2007vj}
  L.~Wulff,
  hep-th/0701129.

\bibitem{Hashimoto:1997gm}
  A.~Hashimoto and W.~Taylor,
  Nucl.\ Phys.\ B {\bf 503} (1997) 193
  [hep-th/9703217].

\bibitem{Sevrin:2001ha}
  A.~Sevrin, J.~Troost and W.~Troost,
  Nucl.\ Phys.\ B {\bf 603} (2001) 389.
  [hep-th/0101192].

\bibitem{Denef:2000rj}
F.~Denef, A.~Sevrin and J.~Troost, 
  Nucl.\ Phys.\ B {\bf 581} (2000) 135
  [hep-th/0002180].



\bibitem{Peeters:2006iu}
  K.~Peeters, J.~Sonnenschein, M.~Zamaklar,
  Phys.\ Rev.\  {\bf D74 } (2006)  106008.
  [hep-th/0606195].

\bibitem{Peeters:2007ab}
  K.~Peeters and M.~Zamaklar,
  Eur.\ Phys.\ J.\ ST {\bf 152} (2007) 113
  [arXiv:0708.1502 [hep-ph]].

\bibitem{Bali:1992ru}
G.~S.~Bali and K.~Schilling,
  Phys.\ Rev.\ D {\bf 47} (1993) 661
  [hep-lat/9208028].

\bibitem{Sommer:1993ce}
R.~Sommer,
  Nucl.\ Phys.\ B {\bf 411} (1994) 839
  [hep-lat/9310022].



\bibitem{Mizher:2011wdb}
A.~J.~Mizher, \emph{private communication}.

\bibitem{Obukhov:1984xb}
I.~A.~Obukhov, V.~K.~Peres-Fernandes, I.~M.~Ternov, V.~R.~Khalilov, Theor.\ Math.\ Phys.\  {\bf 55} (1983) 536  [Teor.\ Mat.\ Fiz.\  {\bf 55} (1983) 335].

\bibitem{Tsai:1972iq}
  W.~-Y.~Tsai and A.~Yildiz,
  Phys.\ Rev.\ D {\bf 4} (1971) 3643.


\bibitem{Djukanovic:2005ag}
  D.~Djukanovic, M.~R.~Schindler, J.~Gegelia and S.~Scherer,
  Phys.\ Rev.\ Lett.\  {\bf 95} (2005) 012001.
  [hep-ph/0505180].



\bibitem{Aharony:2007uu}
  O.~Aharony, K.~Peeters, J.~Sonnenschein, M.~Zamaklar,
  JHEP {\bf 0802 } (2008)  071.
  [arXiv:0709.3948 [hep-th]].

\bibitem{Bergman:2008qv}
  O.~Bergman, G.~Lifschytz, M.~Lippert,
  Phys.\ Rev.\  {\bf D79 } (2009)  105024.
  [arXiv:0806.0366 [hep-th]].

\bibitem{Chernodub:2011tv}
 M.~N.~Chernodub, PoS(FacesQCD) (2010)  021. [arXiv:1104.4404 [hep-ph]].




\bibitem{nele3}
N.~Callebaut, D.~Dudal, \emph{in preparation}.

\bibitem{nele2}
N.~Callebaut, D.~Dudal, \emph{in preparation}.





\end{thebibliography}
\end{document}